\documentclass[superscriptaddress,showpacs,twocolumn,nofootinbib]{revtex4-2}

\usepackage[normalem]{ulem}
\usepackage{graphicx}
\usepackage{color}
\usepackage{verbatim}
\usepackage{subfigure}
\usepackage{amssymb}
\usepackage{amsmath}
\usepackage{comment}
\usepackage{chngcntr}
\usepackage{appendix}
\usepackage{lipsum}
\usepackage{physics}
\usepackage{mathtools}
\usepackage[
colorlinks=true,
citecolor=MidnightBlue,
linkcolor=BrickRed,
urlcolor=black,
breaklinks=true,
]{hyperref}  
\usepackage[
usenames,
dvipsnames
]{xcolor}				

\newcommand{\bq}[0]{\bm{q}}
\newcommand{\ibq}[0]{\text{i}\bm{q}}

\newcommand{\hv}{\hat{v}}
\newcommand{\hatm}{\hat{m}}

\usepackage{amsthm}
\usepackage{bm}
\usepackage[hang,flushmargin]{footmisc}

\begin{document}

\title{The Flux Hypothesis for Odd Transport Phenomena}
\author{Cory Hargus}
\affiliation{Department of Chemical and Biomolecular Engineering, University of California, Berkeley, California 94720, USA}
\affiliation{Université Paris Cité, Laboratoire Matière et Systèmes Complexes (MSC), UMR 7057 CNRS,
F-75205, 75205 Paris, France}

\author{Alhad Deshpande}
\affiliation{Department of Chemical and Biomolecular Engineering, University of California, Berkeley, California 94720, USA}

\author{Ahmad K.\ Omar}
\affiliation{Department of Materials Science and Engineering, University of California, Berkeley, California 94720, USA}
\affiliation{Materials Sciences Division, Lawrence Berkeley National Laboratory, Berkeley, California 94720, USA}

\author{Kranthi K. Mandadapu}
\affiliation{Department of Chemical and Biomolecular Engineering, University of California, Berkeley, California 94720, USA}
\affiliation{Chemical Sciences Division, Lawrence Berkeley National Laboratory, Berkeley, California 94720, USA}

\begin{abstract}
Onsager's regression hypothesis makes a fundamental connection between macroscopic transport phenomena and the average relaxation of spontaneous microscopic fluctuations.
This relaxation, however, is agnostic to odd transport phenomena, in which fluxes run orthogonal to the gradients driving them.
To account for odd transport, we generalize the regression hypothesis, postulating that macroscopic linear constitutive laws 
are, on average, obeyed by microscopic fluctuations, whether they contribute to relaxation or not.
From this ``flux hypothesis,'' Green-Kubo and reciprocal relations follow, elucidating the separate roles of broken time-reversal and parity symmetries underlying various odd transport coefficients.
As an application, we derive and verify the Green-Kubo relation for odd collective diffusion in chiral active matter, first in an analytically-tractable model and subsequently through molecular dynamics simulations of concentrated active spinners.
\end{abstract}

\maketitle

\vspace{0.1in}
\noindent\textbf{\textit{Introduction.}} Ongoing efforts in statistical physics have sought to generalize the canonical theories of phase transitions, rare events, and transport processes to include phenomena which are unique to nonequilibrium systems.
A radical example is odd transport phenomena---such as odd viscosity~\cite{avron1995viscosity,Avr98,banerjee2017odd,liao2019mechanism,Epstein2020,Hargus2020,Han2021,Fruchart2022,hosakaLorentzReciprocalTheorem2023}, odd diffusion~\cite{Hargus2021,Kalz2022,Muzzeddu2022}, and odd thermal conduction~\cite{Fruchart2022oddideal}---in which the gradient of a conserved quantity (momentum, mass, and energy, respectively), drives a corresponding flux in the orthogonal direction.
Recently, odd transport has been shown to exist in a broad class of systems known as ``chiral active matter.''
These systems, which may be biological~\cite{Diluzio2005,friedrichChemotaxisSpermCells2007,Drescher2009,petroffFastMovingBacteriaSelfOrganize2015,tanOddDynamicsLiving2022} or synthetic~\cite{tsai2005chiral,Kummel2013,Nourhani2016,Soni2019,VegaReyes2023}, are characterized by microscopic dynamics breaking time-reversibility and parity symmetry.

The connection between microscopic dynamics and macroscopic transport phenomena was developed in the pioneering works of Onsager, Kubo, Casimir, and Zwanzig, among others~\cite{Onsager1931a,Onsager1931b,Casimir1945,Kubo1957,Kubo1957b, Zwanzig1964, Zwanzig1965}.
By assuming that microscopic fluctuations about equilibrium tend to relax according to the same conservation laws governing macroscopic transport (the \textit{regression hypothesis}), Onsager derived his celebrated \textit{reciprocal relations}, showing that time-reversal symmetry of the microscopic dynamics ensures certain macroscopic symmetries between transport coefficients~\cite{Onsager1931a,Onsager1931b}.
Building upon Onsager's regression hypothesis, Kubo, Yokota, and Nakajima derived \textit{Green-Kubo relations}~\cite{Kubo1957,Kubo1957b} giving a quantitative prediction of these transport coefficients from equilibrium correlations.

Recently, the regression hypothesis was applied beyond its original equilibrium context to derive Green-Kubo relations for odd viscosity in chiral active matter~\cite{Epstein2020,Hargus2020}.
Others have used nonequilibrium linear response theory~\cite{Baiesi2013, Han2021, Poggioli2022} to study the odd response of active systems to external driving.
In general, however, odd fluxes driven by spatial gradients do not fit neatly into either approach: if they do not affect the relaxation of gradients, these fluxes elude the regression hypothesis, yet beyond equilibrium they need not maintain a connection (e.g. Einstein relation) to those driven by external perturbations.
Thus, certain fundamental questions remain.
Namely, how do odd transport phenomena generically arise out of chiral microscopic fluctuations, and when is the breaking of time-reversal symmetry a necessary condition for odd transport?

In this article we develop a general theoretical framework illuminating the microscopic origins of odd transport phenomena and, moreover, revealing cases where odd transport may be possible without breaking time-reversal symmetry.
By introducing the \textit{flux hypothesis}, a reformulation of Onsager's regression hypothesis, we develop general Green-Kubo relations and reciprocal relations, which we apply and validate in the context of odd collective diffusion of chiral active spinners.

\vspace{0.1in}
\noindent\textbf{\textit{Linear constitutive behavior and Onsager's regression hypothesis.}}
Transport phenomena occur when a system is perturbed from its resting state, which may be thermodynamic equilibrium or, more generally, a nonequilibrium steady state.
In this work we consider linear constitutive behavior in which a flux $\bm{J}_i$ is driven by
spatial gradients of conserved quantities $A_j$:
\begin{equation}\label{eq:constitutive}
    \bm{J}_i = -\sum_{j=1}^m \mathbf{M}_{ij} \cdot \bm{\nabla}A_j\,.
\end{equation}
Here, subscripts index over the $m$ quantities being transported (mass, momentum, heat, etc.) such that $\mathbf{M}_{ij} \in \mathbb{R}^{d \times d}$ is one of $m^2$ matrices of constant transport coefficients in $d$ dimensions.
Equation~\eqref{eq:constitutive} encapsulates the constitutive laws of Fick, Fourier, and Newton~\cite{Groot1984,deenAnalysisTransportPhenomena1998,Bird2007} as terms with $i=j$.
Conversely, $i \ne j$ corresponds to cross-couplings, for instance the Soret and Dufour effects~\cite{Groot1984}, in which temperature gradients drive mass flux and mass gradients drive heat flux, respectively.

In the Fourier representation defined by $y^{\bq}(t) = \int d\bm{r}\ e^{-\text{i}\bq\cdot \bm{r}} y(\bm{r}, t)$, Eq.~\eqref{eq:constitutive} becomes
\begin{equation}\label{eq:constitutive-fourier}
\bm{J}_i^{\bq} = -\sum_{j=1}^m \mathbf{M}_{ij} \cdot \ibq A_j^{\bq}\,,
\end{equation}
with wave vector $\bq$.
In the absence of sources or sinks, $A_i^{\bq}$ changes value only as it is transported by the flux $\bm{J}_i ^{\bq}$, and thus obeys the conservation law
\begin{equation}\label{eq:conservation-fourier}
    \frac{d}{d t}A_i^{\bq} = -\ibq \cdot \bm{J}_i ^{\bq}\,.
\end{equation}
Combining Eqs.~\eqref{eq:constitutive-fourier} and~\eqref{eq:conservation-fourier} yields the relaxation equation
\begin{equation}
    \label{eq:relaxation-fourier}
    \frac{d}{d t}A_i^{\bq} = -\sum_{j=1}^m (\bq \cdot \mathbf{M}_{ij}^{\mathrm{even}} \cdot \bq) A_j ^{\bq}\,.
\end{equation}
We have explicitly labeled the \textit{even} part of the transport matrix, defined here as $\mathbf{M}_{ij}^{\mathrm{even}} =\frac{1}{2}(\mathbf{M}_{ij}+\mathbf{M}_{ij}^\intercal)$,
where the transpose $(\cdot)^\intercal$ acts on the spatial elements but not the transport couplings (i.e. subscripts).
In contrast, the \textit{odd} part $\mathbf{M}^\mathrm{odd}_{ij}=\frac{1}{2}(\mathbf{M}_{ij}-\mathbf{M}_{ij}^\intercal)$ drives fluxes perpendicular to the gradient in Eq.~\eqref{eq:constitutive},
and consequently does not enter\footnote{In Newtonian fluids, the stress $\bm{\sigma}$ and strain $\nabla \bm{v}$ are connected by a rank-four viscosity tensor $\eta_{ijkl}$ as $\sigma_{ij} = \eta_{ijkl} \partial_l v_k$, where all subscripts index spatial dimensions. The odd viscosity is usually defined as $\eta_{ijkl}^\mathrm{odd} = \frac{1}{2}(\eta_{ijkl} - \eta_{klij})$, and thus only the part for which $i = k$ generates non-relaxational (i.e. divergence-free) fluxes, while the remaining parts may affect bulk relaxation. See Appendix E for application of the flux hypothesis to odd viscosity. 
} into Eq.~\eqref{eq:relaxation-fourier}.
Nevertheless, $\mathbf{M}^\mathrm{odd}_{ij}$ can affect the evolution of $A_i^{\bq}(t)$ through boundary conditions involving the fluxes $\bm{J}_i^{\bq}$, with significant and sometimes unexpected consequences~\cite{ganeshan2017odd,Hargus2021,Kirkinis2023, batton2023microscopic,louOddViscosityinducedHalllike2022,Ghimenti2023}.

\begin{figure}[t]
    \centering
    \includegraphics[width=.5\textwidth]{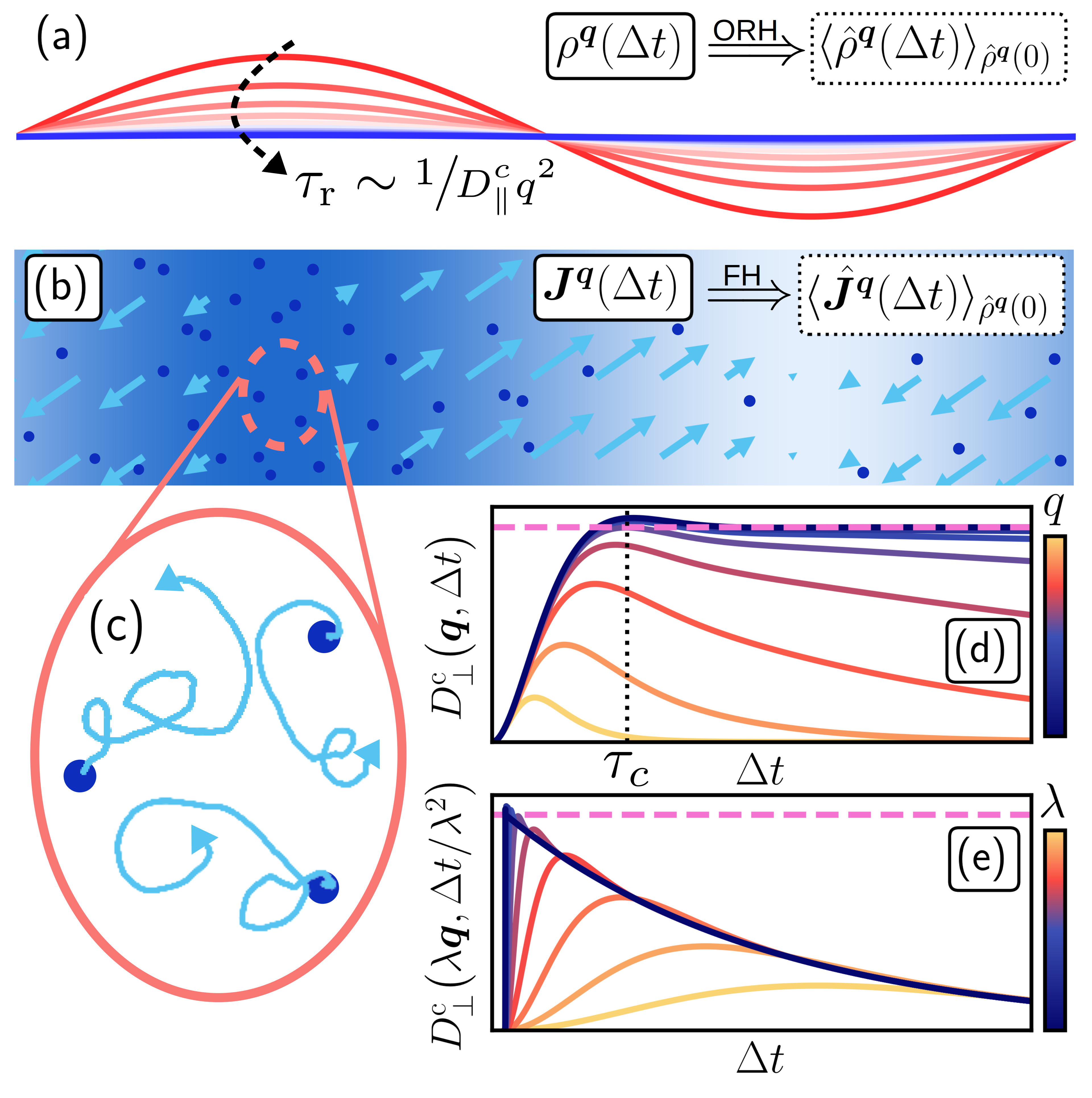}
    \caption{
    Odd collective diffusion.
    \textbf{(a)} An excited density mode $\rho^{\bq}$ of wavelength $1/|\bq|$ relaxes on the timescale $\tau_r \sim 1/(D^\mathrm{c}_\parallel q^2 )$.
    The ORH states that the macroscopic relaxation equation for the density (solid box) is obeyed on average by microscopic fluctuations (dotted box).
    \textbf{(b)} Underlying the relaxation process, the density is transported by fluxes.
    These may include an odd part which eludes the bulk relaxation (vertical component of the light blue arrows).
    The FH states that the macroscopic constitutive law for the flux (solid box), is obeyed on average by microscopic fluctuations (dotted box), for both the even and odd components.
    \textbf{(c)} Typical trajectories of the  chiral Langevin model system described by Eq.~\eqref{eq:chiral-langevin}.
    \textbf{(d)} The odd diffusivity $D_\perp^\mathrm{c}$ of this system (dashed pink line) is obtained on intermediate timescales at small $\bq$ from the analytical solution to the Green-Kubo relations in Eq.~\eqref{eq:2d-langevin-exact-solution}.
    \textbf{(e)} Rescaling by $\lambda$ collapses all curves and recovers Markovian behavior as $\lambda \rightarrow 0$, motivating Eq.~\eqref{eq:Green-Kubo-limit}.
    }
    \label{fig:fig1-assembly}
\end{figure}

To connect $\mathbf{M}^\mathrm{even}_{ij}$ to microscopic dynamics, Onsager~\cite{Onsager1931a,Onsager1931b}
proposed that Eq.~\eqref{eq:relaxation-fourier} applies not only to macroscopic transport, but also to the average behavior of spontaneous microscopic fluctuations about a resting state, as illustrated in Fig.~\ref{fig:fig1-assembly}(a).
This is the Onsager Regression Hypothesis (ORH), expressed mathematically as
\begin{align}
    \begin{split}
    \label{eq:ORH-condition}
    \frac{\langle \hat{A}_i^{\bq}(\Delta t) \rangle_{\{\hat{A}_j^{\bq}(0) = a_j^{\bq}\}} - a_i^{\bq}}{\Delta t}
    = -\sum_{j=1}^m (\bq \cdot \mathbf{M}_{ij}^{\mathrm{even}} \cdot \bq)  a_j^{\bq} \,,
    \end{split}
\end{align}
where $\hat{A}_i^{\bq}$ indicates a fluctuating microscopic quantity corresponding to $A_i^{\bq}$, as obtained for example through coarse-graining of the microscopic dynamics~\cite{Irv50, hardy1982formulas}.
The notation $\langle\cdot\rangle_{\{\hat{A}_j^{\bq}(0) = a_j^{\bq}\}}$ indicates an average over the subensemble compatible with the prescribed initial values $\{a_j^{\bq}\}$.
A more practical statement of the ORH involving correlation functions is obtained by multiplying Eq.~\eqref{eq:ORH-condition} by $a_k^{-\bq}$ and averaging
$\{a_i^{\bq}\}$  over the steady-state distribution of $\{\hat{A}_i^{\bq}\}$, yielding
\begin{align}
    \begin{split}
    \label{eq:ORH-correlators}
    \frac{1}{\Delta t} \big[ \langle \hat{A}_i^{\bq}(\Delta t) \hat{A}_k^{-\bq}(0) \rangle - \langle \hat{A}_i^{\bq}(0) \hat{A}_k^{-\bq}(0) \rangle \big] \\
    = -\sum_{j=1}^m (\bq \cdot \mathbf{M}_{ij}^{\mathrm{even}} \cdot \bq) \langle \hat{A}_j^{\bq}(0) \hat{A}_k^{-\bq}(0) \rangle\,.
    \end{split}
\end{align}

The left-hand sides of Eqs.~\eqref{eq:ORH-condition} and \eqref{eq:ORH-correlators} are formulated as finite-difference time derivatives, valid for $\Delta t$ much shorter than the
fastest macroscopic relaxation time $\tau_r  \sim \min_{ij} 1/|\bq\cdot\mathbf{M}_{ij}^{\mathrm{even}}\cdot\bq|$. 
For Eq.~\eqref{eq:ORH-correlators} to reproduce the Markovian behavior of Eq.~\eqref{eq:relaxation-fourier}, however, $\Delta t$ must also be much longer than the correlation time $\tau_c$, above which the increments
$\hat{A}_i^{\bq}(t + \Delta t) - \hat{A}_i^{\bq}(t)$ and $\hat{A}_j^{\bq}(t) - \hat{A}_j^{\bq}(t - \Delta t)$
can be considered statistically independent for all $i$ and $j$.
Thus $\Delta t$ must satisfy the separation of timescales
\begin{equation}\label{eq:separation-of-timescales}
\tau_c \ll \Delta t \ll \tau_r\,,
\end{equation}
with the upper inequality satisfied in the macroscopic limit $\bq \rightarrow \bm{0}$.
With these considerations, Kubo, Yokota, and Nakajima~\cite{Kubo1957b} derived from Eq.~\eqref{eq:ORH-correlators} the Green-Kubo relations
\begin{equation}\label{eq:GK-1-even}
    \int_0^{\Delta t} dt\ \big\langle \hat{\bm{J}}_i^{\bq}(t) \hat{\bm{J}}^{-\bq}_k(0) \big\rangle^{\mathrm{even}}
    = \sum_{j=1}^m \mathbf{M}_{ij}^{\mathrm{even}} \langle \hat{A}_j^{\bq}(0) \hat{A}^{-\bq}_k(0) \rangle \,,
\end{equation}
where $\hat{\bm{J}}_i^{\bq}$ is the microscopic flux transporting $\hat{A}_i^{\bq}$.

\vspace{0.1in}
\noindent\textbf{\textit{Flux hypothesis.}}
As seen in the previous section, the ORH cannot, by construction, yield any microscopic information about $\mathbf{M}_{ij}^{\mathrm{odd}}$.
Note, however, that the core idea
behind the ORH can in fact be decomposed into two separate assertions.
Firstly, fluctuations $\hat{A}_i^{\bq}$ are conserved and thus obey $\frac{d}{d t}\hat{A}_i^{\bq} = -\ibq \cdot \hat{\bm{J}}_i^{\bq}$ \textit{exactly} for any realization of the fluctuating dynamics, just as for the macroscopic fields in Eq.~\eqref{eq:conservation-fourier}.
Secondly, fluctuations \textit{on average} reproduce macroscopic constitutive behavior.
This suggests that, to access $\mathbf{M}_{ij}^{\mathrm{odd}}$, the latter statement should be posed independently of the former.
Namely, we postulate a flux hypothesis (FH) stating that Eq.~\eqref{eq:constitutive-fourier} applies not only to macroscopic fluxes arising in response to externally-driven gradients, but also to the average behavior of microscopic fluxes arising in response to the spontaneous fluctuations of $\hat{A}_i^{\bq}$.
More precisely,
\begin{equation}~\label{eq:flux-hypothesis-condition}
    \langle\hat{\bm{J}}^{\bq}_i(\Delta t)\rangle_{\{\hat{A}_j^{\bq}(0) = a_j^{\bq}\}} =-\sum_{j=1}^m \mathbf{M}_{ij}\cdot \ibq\, a_j^{\bq} \,,
\end{equation}
as illustrated in Fig.~\ref{fig:fig1-assembly}(b).
Again, $\Delta t$ must satisfy the separation of timescales in Eq.~\eqref{eq:separation-of-timescales}: it should be long enough that constitutive behavior is Markovian, yet short enough that no relaxation of $\hat{A}_j^{\bq}$ has occurred.
Multiplying by $a_k^{-\bq}$ and averaging, as in Eq.~\eqref{eq:ORH-correlators},  yields
\begin{equation}\label{eq:flux-hypothesis-correlators}
        \langle \hat{\bm{J}}_i^{\bq}(\Delta t) \hat{A}_k^{-\bq}(0) \rangle
        = -\sum_{j=1}^m \mathbf{M}_{ij} \cdot \ibq \langle \hat{A}_j^{\bq}(0) \hat{A}_k^{-\bq}(0) \rangle \,.
\end{equation}
Note that contracting both sides of this equation with $\ibq$ and invoking the conservation law recovers Eq.~\eqref{eq:ORH-correlators}.
The ORH is therefore a corollary of the FH.

Assuming stationarity, Eq.~\eqref{eq:flux-hypothesis-correlators} then implies
\begin{equation}\label{eq:GK-1-all}
    \int_0^{\Delta t} dt\ \langle \hat{\bm{J}}_i^{\bq}(t) \hat{\bm{J}}_k^{-\bq}(0) \rangle
    = \sum_{j=1}^m \mathbf{M}_{ij} \langle \hat{A}_j^{\bq}(0) \hat{A}_k^{-\bq}(0) \rangle \,,
\end{equation}
as obtained in Appendix A. Note that Eq.~\eqref{eq:GK-1-all} is identical to Eq.~\eqref{eq:GK-1-even}, except that it holds for all components of $\mathbf{M}_{ij}$, both even and odd.
Finally, defining
\begin{align}
    \label{eq:g-of-q}
    \big[\mathbf{g}^{-1}(\bq)\big]_{ij}  &= \langle \hat{A}_i^{\bq}(0) \hat{A}_j^{-\bq}(0) \rangle\,,  \\
    \label{eq:L-of-q}
    \mathbf{L}_{ij}(\bq, \Delta t) &= \int_0^{\Delta t} dt\ \langle \hat{\bm{J}}_i^{\bq}(t) \hat{\bm{J}}_j^{-\bq}(0) \rangle\,,
\end{align}
and rearranging Eq.~\eqref{eq:GK-1-all} yields the Green-Kubo relations
\begin{equation}\label{eq:Green-Kubo}
\mathbf{M}_{ij}(\bq, \Delta t) = \sum_{k=1}^m \mathbf{L}_{ik}(\bq, \Delta t) g_{kj}(\bq)\,,
\end{equation}
where $\mathbf{L}_{ik}$ encodes the dynamical response and $g_{kj}$ sets the scale of static fluctuations.

\vspace{0.1in}
\noindent\textbf{\textit{Separation of timescales and self-similarity.}}
For both even and odd transport, there remains the ambiguity of identifying an intermediate timescale $\Delta t$ satisfying the inequalities in Eq.~\eqref{eq:separation-of-timescales}.
One possibility is to take first the macroscopic limit $\bm{q} \rightarrow \bm{0}$ and then send $\Delta t \rightarrow \infty$~\cite{kuboStatisticalPhysicsII1991,EvansMorris}.
Moreover, for gradient-driven transport, self-similarity of the time correlation functions suggests rescaling space and time as $\bq \mapsto \lambda \bq$ and $\Delta t \mapsto \Delta t / \lambda^2$ in $\mathbf{M}_{ij}(\bq, \Delta t)$, as illustrated in Figs.~\ref{fig:fig1-assembly}(d,e). In this way, the macroscopic and Markovian characteristics of Eq.~\eqref{eq:constitutive-fourier} are both recovered at small values of the scaling parameter $\lambda$, in line with work by Zwanzig~\cite{Zwanzig1964}.
The Green-Kubo relations in Eq.~\eqref{eq:Green-Kubo} may then be evaluated as
\begin{equation}\label{eq:Green-Kubo-limit}
\mathbf{M}_{ij} = \lim_{\Delta t \rightarrow 0} \lim_{\lambda \rightarrow 0} \sum_{k=1}^m \mathbf{L}_{ik}(\lambda \bq, \Delta t/\lambda^2) g_{kj}(\lambda \bq) \,,
\end{equation}
noting that the limits do not commute.
The inner limit recovers the Markovian, macroscopic phenomenology of Eq.~\eqref{eq:constitutive}.
The outer limit then ensures that correlations are instantaneous with respect to the relaxation time.

\vspace{0.1in}
\noindent\textbf{\textit{Reciprocal relations.}}
By connecting fluctuations to transport phenomena, the Green-Kubo relations~\eqref{eq:Green-Kubo-limit} reveal the microscopic symmetries involved in odd transport.
Most fundamentally, odd transport requires antisymmetry at the level of the time correlation functions in Eq.~\eqref{eq:L-of-q}.
For dynamics obeying time-reversal symmetry (TRS), this implies the reciprocal relations
\begin{equation}
    \mathbf{L}_{ij} 
    \stackrel{\text{TRS}}{=} 
    h_i h_j\mathbf{L}_{ji}^\intercal\ ,
\end{equation}
where $h_{i}\in\{-1,1\}$ indicates whether the flux $\hat{\bm{J}}_i^{\bq}$ is an even ($h_i=1$) or odd ($h_i=-1$) function of the particle velocities.
For instance, a fluid stress will be even while a diffusive flux is odd.
Onsager's regression hypothesis leads to $\mathbf{L}_{ij}^\mathrm{even} \stackrel{\text{TRS}}{=} h_i h_j\mathbf{L}_{ji}^\mathrm{even}$, while the flux hypothesis yields the additional odd reciprocal relations
\begin{equation}\label{eq:odd-reciprocal}
    \mathbf{L}_{ij}^\mathrm{odd}
    \stackrel{\text{TRS}}{=}
    -h_i h_j\mathbf{L}_{ji}^\mathrm{odd}\,.
\end{equation}
When $i=j$, we find $\mathbf{L}_{ii}^\mathrm{odd} \stackrel{\text{TRS}}{=} -\mathbf{L}_{ii}^\mathrm{odd} = 0$, indicating that TRS-breaking is indeed a requirement for self-induced odd transport. 
Interestingly, odd cross terms ($i \ne j$) are in principal possible without breaking TRS, requiring only parity symmetry breaking.
This suggests the possibility of equilibrium (i.e. passive) matter exhibiting odd cross couplings, for instance between viscous and diffusive transport.

\vspace{0.1in}
\noindent\textbf{\textit{Application to odd collective diffusion.}}
\begin{figure*}[t!]
    \centering
    \includegraphics[width=\textwidth]{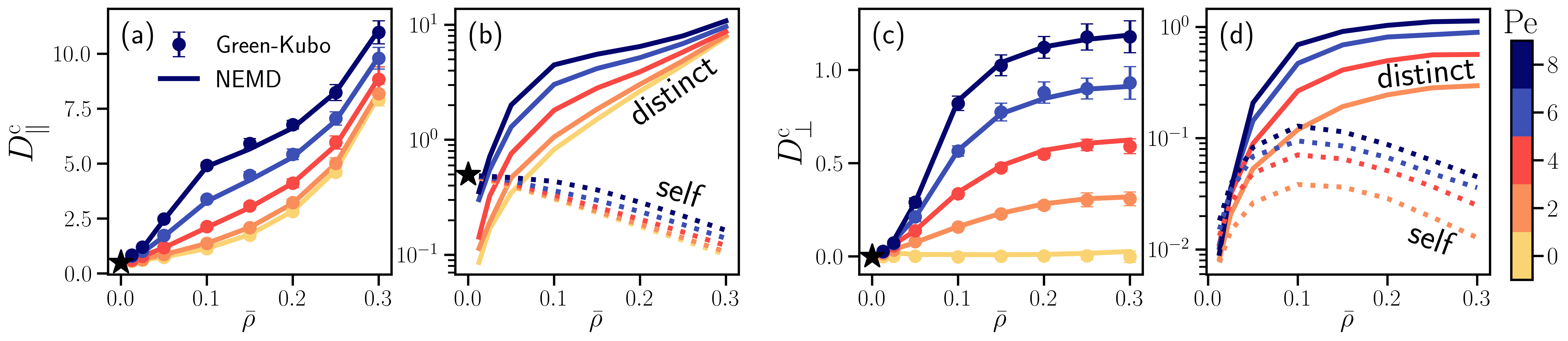}
    \caption{
    Collective diffusion of concentrated active spinners. \textbf{(a, c)} Green-Kubo estimates of $D_\parallel^\mathrm{c}$ and $D_\perp^\mathrm{c}$ from Eq.~\eqref{eq:GK-collective-diffusion-2} (error bars for 95 \% confidence intervals) and NEMD direct measurements from Eq.~\eqref{eq:NEMD} (solid lines, error bars smaller than the symbol size).
    In the dilute limit, $D_\parallel^\mathrm{c} = \frac{1}{2}$ and $D_\perp^\mathrm{c}=0$, marked by black stars.
    \textbf{(b, d)}. Contributions of $\mathbf{D}^\mathrm{self}$ (dotted) and \ $\mathbf{D}^\mathrm{distinct}$ (solid), showing the latter to dominate in concentrated solutions ($\bar{\rho} \gtrsim 0.1$).
    }
    \label{fig:dumbbell-results}
\end{figure*}
We now apply the preceding theoretical framework in a concrete and novel setting involving odd collective diffusion.
Application to other odd transport phenomena proceeds analogously, as shown in Appendix E for odd thermal conductivity and odd viscosity\footnote{Note also that while the ORH recovers a set of six Green-Kubo relations for the eight viscosity coefficients involved in isotropic, compressible two-dimensional chiral fluids~\cite{Epstein2020}, the FH offers the advantage of recovering eight independent relations for each of the viscosity coefficients.}.
Collective (or ``down-gradient'') diffusion in concentrated solutions, whether passive or active, involves the flux $\bm{J}$ of solute relative to a stationary medium driven by gradients in its density $\rho$, with the linear constitutive law
\begin{equation}\label{eq:ficks-law}
\bm{J}(\bm{r}, t) = - \mathbf{D}^{\mathrm{c}}\big(\rho\big) \cdot \bm{\nabla} \rho(\bm{r},t)\,,
\end{equation}
as depicted in Fig.~\ref{fig:fig1-assembly}.
Unlike self-diffusion in dilute solutions, whose odd part has been described previously~\cite{Hargus2021,Kalz2022,Muzzeddu2022}, the collective diffusivity $\mathbf{D}^{\mathrm{c}}$ depends on the solute density due to interactions between solute particles.
In two-dimensional isotropic systems, $\mathbf{D}^\mathrm{c} = D^\mathrm{c}_{\parallel} \boldsymbol\delta - D^\mathrm{c}_{\perp} \boldsymbol{\epsilon}$,
where \mbox{$\bm{\delta} = \begin{bmatrix} 1 & 0\\ 0 & 1 \end{bmatrix}$ and
$\bm{\epsilon} = \begin{bmatrix} 0 & 1\\ \text{-}1 & 0 \end{bmatrix}$} and
$D^\mathrm{c}_{\parallel} $ and $D^\mathrm{c}_{\perp}$ are the even and odd collective diffusivity, respectively.
For small perturbations about a homogeneous density, $\rho(\bm{r}, t) = \bar{\rho} +  \delta \rho(\bm{r}, t)$, Eq.~\eqref{eq:ficks-law} assumes the linear form 
\begin{equation}\label{eq:ficks-law-linearized}
    \bm{J}(\bm{r}, t) = - \mathbf{D}^{\mathrm{c}}(\bar{\rho}) \cdot \bm{\nabla} \delta \rho(\bm{r},t)\,,
\end{equation}
matching Eq.~\eqref{eq:constitutive}, with $\mathbf{M} = \mathbf{D}^\mathrm{c}(\bar{\rho})$ and $A = \delta \rho$.

The Green-Kubo relations in Eq.~\eqref{eq:Green-Kubo} are then
\begin{align}\label{eq:GK-collective-diffusion}
        &\mathbf{D}^{\mathrm{c}}(\bq, \Delta t\,;\,\bar{\rho}) = \frac{\int_0^{\Delta t}dt\ \langle\hat{\bm{J}}^{\bq}(t)\hat{\bm{J}}^{-\bq}(0)\rangle}{\langle \hat{\rho}^{\bq}\hat{\rho}^{-\bq}\rangle}\,,
\end{align}
identifying $g^{-1}(\bq) = \langle \hat{\rho}^{\bq}\hat{\rho}^{-\bq}\rangle$ with the static structure factor and $\mathbf{L}(\bq, \Delta t) = \int_0^{\Delta t}dt\ \langle\hat{\bm{J}}^{\bq}(t)\hat{\bm{J}}^{-\bq}(0)\rangle$ as the solute flux autocorrelation function (see Appendix B.1).
For $N$ solute particles at positions $\bm{r}_{\alpha}$ with velocities $\bm{v}_{\alpha}$, we define the microscopic fields $\hat{\rho}(\bm{r}, t) = \sum_{\alpha=1}^N \delta\big(\bm{r} - \bm{r}_\alpha(t)\big)$, where $\langle \hat{\rho} \rangle = \bar{\rho}$, and $\hat{\bm{J}}(\bm{r}, t) = \sum_{\alpha=1}^N \bm{v}_\alpha(t) \delta\big(\bm{r} - \bm{r}_\alpha(t)\big)$.
Equation~\eqref{eq:GK-collective-diffusion} is then
\begin{equation}\label{eq:GK-collective-diffusion-2}
        \mathbf{D}^{\mathrm{c}}(\bq, \Delta t\,;\,\bar{\rho})
        = \frac{\sum\limits_{\alpha \beta} \int_0^{\Delta t} dt \langle \bm{v}_\alpha(t) \bm{v}_\beta(0) \cos\big(\bm{q} \cdot \bm{\Delta r}_{\alpha \beta}(t) \big)\rangle}
        {\sum\limits_{\alpha \beta} \langle \cos\big(\bq \cdot \bm{\Delta r}_{\alpha \beta}(0) \big)\rangle } \,,
\end{equation}
having defined $ \bm{\Delta r}_{\alpha \beta}(t) := \bm{r}_\alpha(t) - \bm{r}_\beta(0)$.

In the dilute limit, correlations vanish for $\alpha \ne \beta$.
For identical particles, Eq.~\eqref{eq:GK-collective-diffusion-2} then simplifies to the Green-Kubo relation for self-diffusion of a single particle
\begin{align}\label{eq:self-diffusion}
        \mathbf{D}^\mathrm{self} &= \lim_{\Delta t\rightarrow 0}\lim_{\lambda\rightarrow 0}\int_0^{\Delta t/\lambda^2}\hspace{-6mm} dt\langle \bm{v}(t)\bm{v}(0)\cos\big(\lambda\bq\cdot\bm{\Delta r}(t)\big)\rangle \notag \\
        &= \int_0^\infty dt\ \langle \bm{v}(t)\bm{v}(0) \rangle\,,
\end{align}
consistent with Ref.~\cite{Hargus2021}, showing $D_\perp^{\mathrm{self}}$ to arise from chiral random motion, as reviewed in Appendix B.2.

\vspace{0.1in}
\noindent\textbf{\textit{Chiral Langevin diffusion.}}
From Eq.~\eqref{eq:odd-reciprocal}, it is clear that any system exhibiting odd diffusion must break TRS.
A straightforward model system meeting this requirement
consists of non-interacting particles described by the Langevin equation
\begin{align}\label{eq:chiral-langevin}
\begin{split}
    \dot{\bm{r}}_\alpha(t) &= \bm{v}_\alpha(t) \\
    \dot{\bm{v}}_\alpha(t) &= -\bm{\gamma} \bm{v}_\alpha(t) + \bm{\eta}_\alpha(t) \,,
\end{split}
\end{align}
with the asymmetric friction $\bm{\gamma} = \gamma_\parallel \boldsymbol{\delta} + \gamma_\perp \boldsymbol{\epsilon}$
and thermal noise obeying
${\langle \bm{\eta}_\alpha(t) \bm{\eta}_\beta(t') \rangle = 2 k_{\mathrm{B}}T \bm{\gamma} \delta_{\alpha \beta} \delta(t-t')}$.
The odd friction $\gamma_\perp$ breaks TRS by inducing chiral motion (see Fig.~\ref{fig:fig1-assembly}(c)), playing an identical role to the Lorentz force affecting particles carrying charge $q$ and moving in the plane normal to a magnetic field $B$, in which case $|\gamma_\perp| = |qB|$.
More generally, odd friction can be a consequence of chiral dynamics intrinsic to the solute or arising out of interactions with the solvent~\cite{Reichhardt2019,Poggioli2022}, and serves as a useful model for detailed-balance-violating currents frequently encountered in living systems~\cite{Dinis2012,gnesottoBrokenDetailedBalance2018,obyrneTimeIrReversibility2021}.

Given the non-interacting nature of the particles, the self-diffusivity then follows directly from Eq.~\eqref{eq:self-diffusion} as $\mathbf{D}^{\mathrm{self}} = k_\mathrm{B} T \bm{\gamma}^{-1}$, with even and odd components
\begin{align}
    {D_\parallel^\mathrm{self}} = k_\mathrm{B}T \frac{\gamma_\parallel}{\gamma_\parallel^2 + \gamma_\perp^2}\,, \hspace{0.1in}
    {D_\perp^\mathrm{self}} = k_\mathrm{B}T \frac{\gamma_\perp}{\gamma_\parallel^2 + \gamma_\perp^2}\, .
\end{align}
Independently, the collective diffusivity $\mathbf{D}^{\mathrm{c}}(\bq, \Delta t)$ can be solved exactly (see Appendix C, \cite{Chandrasekhar1943, VanKampen1981}) yielding
\begin{equation}\label{eq:2d-langevin-exact-solution}
    \mathbf{D}^\mathrm{c}(\bq, \Delta t) = k_\mathrm{B} T \bm{\gamma}^{-1}(\bm{\delta} - e^{-\bm{\gamma}\Delta t}) e^{-\frac{1}{4}q^2 \langle | \bm{\Delta r}(\Delta t)|^2 \rangle}\,.
\end{equation}
Note that $\lim_{\Delta t \rightarrow 0}\mathbf{D}^\mathrm{c}(\bq, \Delta t) = \lim_{\Delta t \rightarrow \infty} \mathbf{D}^\mathrm{c}(\bq, \Delta t) = 0$ for any finite $\bq$.
The correct value of $\mathbf{D}^{\mathrm{c}}$ is attained only on intermediate timescales, as plotted in Fig.~\ref{fig:fig1-assembly}(d), or upon rescaling (Fig.~\ref{fig:fig1-assembly}(e)) and applying the limits of Eq.~\eqref{eq:Green-Kubo-limit} in the correct order, resulting in
\begin{align}
        \mathbf{D}^\mathrm{c} &= \lim_{\Delta t \rightarrow 0} \lim_{\lambda \rightarrow 0} \mathbf{D}^\mathrm{c}(\lambda \bq, \Delta t / \lambda^2)\\
        &= \lim_{\Delta t \rightarrow 0} k_\mathrm{B} T \bm{\gamma}^{-1} e^{- q^2 D_\parallel^\mathrm{self} \Delta t }
        = k_\mathrm{B} T \bm{\gamma}^{-1} \notag\,,
\end{align}
confirming that $\mathbf{D}^\mathrm{c} = \mathbf{D}^{\text{self}}$ for both the even and odd components, due to the non-interacting setting.

\vspace{0.1in}
\noindent\textbf{\textit{Collective diffusion of concentrated active spinners.}}
Odd collective diffusion can also arise out of collisions between spinning solute particles, as encountered in a range of physical contexts~\cite{Diluzio2005,friedrichChemotaxisSpermCells2007,Drescher2009,petroffFastMovingBacteriaSelfOrganize2015,tanOddDynamicsLiving2022,tsai2005chiral,Kummel2013,Soni2019,VegaReyes2022,VegaReyes2023}.
Consider a system of chiral active spinners with positions $\bm{r}_\alpha$ and orientations $\theta_\alpha$, evolving as
\begin{align}\label{eq:spinners-eom}
    \begin{split}
        \dot{\bm{r}}_\alpha &= \bm{v}_\alpha\,, \hspace{6mm} \dot{\theta}_\alpha = \Omega_\alpha\\
        m \dot{\bm{v}}_\alpha &= -\gamma \bm{v}_\alpha - \bm{\nabla}_{\bm{r}_\alpha}(V^{\mathrm{int}} + V^{\mathrm{ext}})  + \bm{\eta}_\alpha \\
        I \dot{\Omega}_\alpha &= -\gamma_\theta \Omega_\alpha + \Gamma^a - \partial_{\theta_\alpha}(V^{\mathrm{int}} + V^{\mathrm{ext}}) + \xi_\alpha\,.
    \end{split}
\end{align}
Here, $m \bm{v}_\alpha(t)$ and $I \Omega_\alpha(t)$ are the linear and angular momenta, with mass $m$, moment of inertia $I$, friction coefficients $\gamma$ and $\gamma_\theta$, and noise obeying the fluctuation-dissipation relations
${\langle \bm{\eta}_\alpha(t) \bm{\eta}_\beta(t') \rangle = 2 k_{\mathrm{B}}T \gamma \delta_{\alpha \beta} \delta(t-t') \bm{\delta}}$ and
${\langle \xi_\alpha(t) \xi_\beta(t') \rangle = 2 k_{\mathrm{B}}T \gamma_\theta \delta_{\alpha \beta} \delta(t-t')}$.
Pairwise interactions depend on the relative position and orientation of the particles, i.e.
$V^{\mathrm{int}} = \frac{1}{2}\sum_{\alpha \ne \beta}V_{\alpha \beta}(|\bm{r}_\alpha-\bm{r}_\beta|, |\theta_\alpha-\theta_\beta|)$.
In practice, we simulate rigid dumbbells using LAMMPS software~\cite{lammps} (see Appendix D for simulation details).

Chirality is imposed on the rotational motion via the active torque $\Gamma^a$, whose magnitude relative to thermal forces is quantified by the Péclet number $\mathrm{Pe} = \Gamma^a / k_{\mathrm{B}}T$.
Chirality then enters the translational motion through $V^{\mathrm{int}}$.
In other words, $\Gamma^a$ biases the interaction statistics, such that the average collision path of two spinners is parity-asymmetric,  though $V^{\mathrm{int}}$ contains no inherent asymmetry~\cite{liao2019mechanism,Fruchart2022oddideal}.
Thus, in contrast to the previous section, odd diffusion is expected to vanish in the dilute limit.

The results of computing $D_\parallel^{\mathrm{c}}$ and $D_\perp^{\mathrm{c}}$ from the Green-Kubo relations in Eq.~\eqref{eq:GK-collective-diffusion} are shown in Figs.~\ref{fig:dumbbell-results}(a,c). Note that the zero-wave vector solute flux $\hat{\bm{J}}^{\bq=\mathbf{0}} = \sum_{\alpha=1}^N \bm{v}_\alpha(t) = N \bar{\bm{v}}(t)$,  where the center-of-mass velocity $\bar{\bm{v}}(t)$, loses all information about pairwise interactions $V^\mathrm{int}$, while density fluctuations also vanish at $\bq = \bm{0}$.
Thus, $\mathbf{D}^\mathrm{c}(\bq, \Delta t)$ can only be evaluated at small but finite $\bq$, i.e. in the limit form of Eq.~\eqref{eq:Green-Kubo-limit}.
To validate the Green-Kubo estimates, we independently calculate $\mathbf{D}^{\mathrm{c}}$ using nonequilibrium molecular dynamics (NEMD), imposing a small perturbation $V^{\mathrm{ext}}(x) = -\delta V \cos( q x)$ and measuring the steady-state density and flux, which admit the solution (see Appendix D)
\begin{align}\label{eq:NEMD}
    \begin{split}
        \rho(x) &= \bar{\rho}\bigg(1 + \frac{\delta V}{D_\parallel^{\mathrm{c}} \gamma} \cos(qx) \bigg)\,,\\
        J_y &= -D_\perp^{\mathrm{c}} \partial_x \rho = \frac{D_\perp^{\mathrm{c}} \bar{\rho} q \delta V}{D_\parallel^{\mathrm{c}} \gamma}\sin(qx)\,.
    \end{split}
\end{align}
Note that the odd flux $J_y$ persists even at steady state.
Figures~\ref{fig:dumbbell-results}(a,c) show these values to be in good agreement with Green-Kubo predictions from
Eq.~\eqref{eq:GK-collective-diffusion-2} across a range of densities $\bar{\rho}$ and active torques $\mathrm{Pe}$.

The Green-Kubo formulas in Eq.~\eqref{eq:GK-collective-diffusion-2} further allow decomposition of the diffusion tensor as $\mathbf{D}^\mathrm{c} = \mathbf{D}^\mathrm{self} + \mathbf{D}^\mathrm{distinct}$, where the ``distinct'' part comprises the $\alpha \ne \beta$ terms in the summation.
These are plotted in Figs.~\ref{fig:dumbbell-results}(b,d), where the distinct contributions are seen to dominate $\mathbf{D}^\mathrm{c}$ for $\bar{\rho} \gtrsim 0.1$.
The self-diffusivity $\mathbf{D}^\mathrm{self}$ diminishes at high density, where the motion of individual spinners is increasingly impeded by collisions.
As anticipated, ${D}^\mathrm{self}_\perp$ also vanishes in the dilute limit, thereby attaining a maximum at intermediate density.

\vspace{0.1in}
\noindent\textbf{\textit{Conclusion.}} Macroscopic transport coefficients are encoded in microscopic, steady-state fluctuations, as formulated in Onsager's regression hypothesis. 
Yet the existence of odd transport phenomena, as observed in an ever-widening range of active systems, necessitates making this connection directly at the level of constitutive behavior, rather than bulk relaxation.
The result is the flux hypothesis, which we have invoked to obtain Green-Kubo relations, which we subsequently validated in an application to odd collective diffusion.
Further, the ensuing reciprocal relations reveal that time-reversal and parity symmetry breaking are requirements for odd self-induced transport, while only the latter symmetry must be broken for odd cross-couplings.
Future directions include further validation of the flux hypothesis in nonequilibrium settings, and applications to odd cross-couplings, in both passive and active systems.

\vspace{0.1in}

\noindent\textbf{\textit{Acknowledgements.}} We are grateful to Lila Sarfati, Frédéric van Wijland, Julien Tailleur, and Karthik Shekhar for stimulating discussions. C.H. and A.D. are supported by the National Science Foundation Graduate Research Fellowship Program under Grant Nos.\ DGE 1752814 and 2146752, respectively. C.H. also acknowledges the support of ANR grant THEMA.
A.K.O. is supported by the Laboratory Directed Research and Development Program of Lawrence Berkeley National Laboratory under U.S. Department of Energy Contract No. DE-AC02-05CH11231.
K.K.M is supported by Director, Office of Science, Office of Basic Energy Sciences, of the U.S. Department of Energy under contract No. DEAC02-05CH11231. 
This research used resources of the National Energy Research Scientific Computing Center (NERSC), a U.S. Department of Energy Office of Science User Facility located at Lawrence Berkeley National Laboratory, using NERSC award BES-ERCAP0023682.

\bibliography{ref}

\onecolumngrid
\renewcommand{\theequation}{\thesection.\arabic{equation}}
\renewcommand{\thefigure}{\thesection.\arabic{figure}}
\renewcommand{\thesection}{\Alph{section}}
\renewcommand{\thesubsection}{\thesection.\arabic{subsection}}

\makeatletter
\renewcommand{\p@subsection}{}  
\renewcommand{\p@subsubsection}{}
\makeatother

\newpage
\begin{center}
\textbf{Appendices} \\
\vspace{0.05in}
\end{center}

\setcounter{equation}{0}
\setcounter{footnote}{0}
\setcounter{section}{0}
\setcounter{figure}{0}

\section{Green-Kubo Relations from the Flux Hypothesis}\label{section:appendix-ORH}

In this Appendix we provide the details of obtaining the Green-Kubo relations (Eq.~\eqref{eq:Green-Kubo} of the main text) from the Flux Hypothesis, starting with Eq.~\eqref{eq:flux-hypothesis-correlators} of the main text:
\begin{equation}\label{eq-SI:flux-hypothesis-correlators}
    \langle \hat{\bm{J}}_i^{\bq}(\Delta t) \hat{A}_k^{-\bq}(0) \rangle
        = -\sum_{j=1}^m \mathbf{M}_{ij} \cdot \ibq \langle \hat{A}_j^{\bq}(0) \hat{A}_k^{-\bq}(0) \rangle \,. 
\end{equation}
Note that the right-hand side of Eq.~\eqref{eq-SI:flux-hypothesis-correlators} preserves all information about $\mathbf{M}_{ij}^{\text{odd}}$, since the equality must hold for all orientations of the wave vector $\bq$.
The left-hand side may be expanded as
\begin{equation}
    \langle \hat{\bm{J}}_i^{\bq}(\Delta t) \hat{A}_k^{-\bq}(0) \rangle =  \langle \hat{\bm{J}}_i^{\bq}(0) \hat{A}_k^{-\bq}(0) \rangle + \int_0^{\Delta t}\frac{d}{d t} \langle \hat{\bm{J}}_i^{\bq}(t) \hat{A}_k^{-\bq}(0) \rangle dt \,.
\end{equation}
In what follows, we assume for simplicity that the static correlator $\langle \hat{\bm{J}}_i^{\bq}(0) \hat{A}_k^{-\bq}(0) \rangle = 0$.
The remaining correlation function can be simplified as follows,
\begin{align}~\label{eq-SI:stationarity}
    \begin{split}
        \frac{d}{dt}\langle \hat{\bm{J}}_i^{\bq}(t) \hat{A}_k^{-\bq}(0) \rangle &= \frac{d}{dt}\langle \hat{\bm{J}}_i^{\bq}(0) \hat{A}_k^{-\bq}(-t) \rangle \\
        = &-\frac{d}{dt'}\langle \hat{\bm{J}}_i^{\bq}(0) \hat{A}_k^{-\bq}(t') \rangle\big|_{t'=-t} \\
        = &-\langle \hat{\bm{J}}_i^{\bq}(0) \hat{\bm{J}}_k^{-\bq}(-t) \rangle \cdot \ibq \\
        = &-\langle \hat{\bm{J}}_i^{\bq}(t) \hat{\bm{J}}_k^{-\bq}(0) \rangle \cdot \ibq\,,
    \end{split}
\end{align}
where we have used the stationarity of the correlation function to transfer the time derivative to the state variable $\hat{A}_k^{-\bq}$ and applied the conservation equation
\begin{equation}~\label{eq-SI:FH-IK}
    \frac{d}{dt}\hat{A}_i^{\bq}(t)=-\ibq\cdot\hat{\bm{J}}^{\bq}_i(t)\,.
\end{equation}
Note that Eq.~\eqref{eq-SI:FH-IK} must hold not only on average but also identically for any realization of the microscopic fluctuating variables \cite{hardy1982formulas}.
Inputting Eq.~\eqref{eq-SI:stationarity} into Eq.~\eqref{eq-SI:flux-hypothesis-correlators} yields
\begin{equation} \label{eq-SI-fh-3}
    \int_0^{\Delta t} dt\ \langle \hat{\bm{J}}_i^{\bq}(t) \hat{\bm{J}}_k^{-\bq}(0) \rangle\cdot \ibq
    = \sum_{j=1}^m \mathbf{M}_{ij} \langle \hat{A}_j^{\bq}(0) \hat{A}_k^{-\bq}(0) \rangle\cdot \ibq \,. 
\end{equation}
The fact that this equality holds for any orientation of $\bq$ implies that
\begin{equation} \label{eq-SI:GK-1-all}
    \int_0^{\Delta t} dt\ \langle \hat{\bm{J}}_i^{\bq}(t) \hat{\bm{J}}_k^{-\bq}(0) \rangle
    = \sum_{j=1}^m \mathbf{M}_{ij} \langle \hat{A}_j^{\bq}(0) \hat{A}_k^{-\bq}(0) \rangle \,.
\end{equation}
Equation~\eqref{eq-SI:GK-1-all} exactly mirrors the Green-Kubo relations obtained from the Onsager regression hypothesis~\eqref{eq:GK-1-even}, except that we have retained all information about $\mathbf{M}_{ij}^{\text{odd}}$. Defining
\begin{equation}\label{eq-SI:g-of-q}
    \big[\mathbf{g}^{-1}(\bq)\big]_{ij}  = \langle \hat{A}_i^{\bq}(0) \hat{A}_j^{-\bq}(0) \rangle 
\end{equation}
and
\begin{equation}\label{eq-SI:L-of-q}
    \mathbf{L}_{ij}(\bq, \Delta t) = \int_0^{\Delta t} dt\ \langle \hat{\bm{J}}_i^{\bq}(t) \hat{\bm{J}}_j^{-\bq}(0) \rangle \,,
\end{equation}
where we note that $\mathbf{g}(\bq)$ is invertible as state variables must be orthogonal, Eq.~\eqref{eq-SI:GK-1-all} then yields a closed form expression for $\mathbf{M}_{ij}(\bq,\Delta t)$:
\begin{equation}\label{eq-SI:Green-Kubo}
    \mathbf{M}_{ij}(\bq,\Delta t) = \sum_{k=1}^m \mathbf{L}_{ik}(\bq, \Delta t) g_{kj}(\bq)  \,,
\end{equation}
recovering Eq.~\eqref{eq:Green-Kubo} of the main text. Obtaining the macroscopic transport coefficient $\mathbf{M}_{ij}$ amounts to taking the macroscopic limit $\bq\rightarrow\bm{0}$ and choosing an appropriate correlation time $\Delta t$ of Eq.~\eqref{eq-SI:Green-Kubo}. In what follows, we work with both Eqs.~\eqref{eq-SI:flux-hypothesis-correlators} and~\eqref{eq-SI:Green-Kubo} when applying the flux hypothesis to collective diffusion.

\section{Green-Kubo Relations for Odd Diffusion}
\subsection{Collective Diffusion}\label{section:appendix-odd-collective-diffusion}
The linear constitutive equation governing the flux of a species $\bm{J}$ in response to density gradients $\rho$ is Fick's Law
\begin{equation}\label{eq-SI:ficks-law}
\bm{J}(\bm{r}, t) = - \mathbf{D}^\mathrm{c}(\rho) \bm{\nabla} \rho(\bm{r},t) \,,
\end{equation}
where $\rho (\bm{r}, t) \in \mathbb{R}$ is the density field, $\bm{J}(\bm{r}, t) \in \mathbb{R}^{d}$ is the flux and $\mathbf{D}^\mathrm{c} \in \mathbb{R}^{d \times d}$ is the diffusivity.
Non-linearity in this constitutive relation is accounted for through the density dependence of $\mathbf{D}^\mathrm{c}(\rho)$.
For small perturbations $\delta \rho(\bm{r}, t)$ about the constant and homogeneous density $\bar{\rho}$, Eq.~\eqref{eq-SI:ficks-law} is linearized as
\begin{equation}\label{eq-SI:ficks-linear}
    \bm{J}(\bm{r}, t) = - \mathbf{D}^\mathrm{c}(\bar{\rho}) \bm{\nabla} \delta \rho(\bm{r},t)\,.
\end{equation}
Introducing the Fourier transform $y^{\bq}(t) = \int d\bm{r}\ e^{-\text{i}\bq\cdot \bm{r}} y(\bm{r}, t)$, Eq.~\eqref{eq-SI:ficks-linear} can be expressed as
\begin{equation}
    \bm{J}^{\bq}(t) = - \mathbf{D}^\mathrm{c}(\bar{\rho}) \cdot \ibq \delta \rho^{\bq}(t)\,,
\end{equation}
where the density modes are $\delta\rho^{\bq} = \rho^{\bq} - \bar{\rho}\delta(\bq)$.
Since we are only interested in Fourier modes with finite $\bq$, we can ignore the $\bar{\rho}\delta(\bq)$ contribution, and replace $\delta \rho^{\bq}(t)$ with $\rho^{\bq}(t)$.

The flux hypothesis for collective diffusion, as stated from Eq.~\eqref{eq:flux-hypothesis-correlators} of the main text, is
 \begin{equation}\label{eq-SI:flux-hypothesis-diffusion}
 \mathbf{D}^\mathrm{c} \cdot \ibq \langle \hat{\rho}^{\bq}\hat{\rho}^{-\bq} \rangle = -\langle \hat{\bm{J}}^{\bq}(\Delta t) \hat{\rho}^{-\bq}(0) \rangle\,.
\end{equation}
Invoking the conservation law $\frac{d}{dt}\hat{\rho}^{\bq} = -\ibq \cdot \hat{\bm{J}}^{\bq}$ together with stationarity, as in Appendix~\ref{section:appendix-ORH}, leads directly to the Green-Kubo relations
\begin{equation}\label{eq-SI:collective-diffusion-green-kubo}
    \mathbf{D}^\mathrm{c}(\bq,\Delta t) = \langle\delta \hat{\rho}^{\bq}\delta \hat{\rho}^{-\bq}\rangle^{-1} \int_0^{\Delta t}dt\langle \hat{\bm{J}}^{\bq}(t)\hat{\bm{J}}^{-\bq}(0)\rangle\,,
\end{equation}
mirroring Eq.~\eqref{eq:Green-Kubo} of the main text. 

For an $N$-particle system with positions $\{\bm{r}_\alpha\}$ and velocities $\{\bm{v}_\alpha\}$, the flux and density fields are defined through coarse-graining~\cite{Irv50,hardy1982formulas} as
\begin{align}
    \label{eq-SI:phase-density}
    \hat{\rho}(\bm{r}, t) &:= \sum_{\alpha=1}^N \delta(\bm{r} - \bm{r}_\alpha)\,, \\
    \label{eq-SI:phase-flux}
    \hat{\bm{J}}(\bm{r}, t) &:= \sum_{\alpha=1}^N \bm{v}_\alpha \delta(\bm{r} - \bm{r}_\alpha)\,,
\end{align}
with corresponding Fourier modes given by 
\begin{align}
    \label{eq-SI:phase-density-fourier}
    \hat{\rho}^{\bq}(t) &= \sum_{\alpha=1}^N e^{-\ibq \cdot \bm{r}_\alpha(t)}\,, \\
    \label{eq-SI:phase-flux-fourier}
    \hat{\bm{J}}^{\bq}(t) &= \sum_{\alpha=1}^N \bm{v}_\alpha(t) e^{-\ibq \cdot \bm{r}_\alpha(t)}\,.
\end{align}
Equation~\eqref{eq-SI:flux-hypothesis-diffusion} is then
\begin{equation}\label{eq-SI:imaginary-correlator-1}
    \mathbf{D}^\mathrm{c} \cdot \ibq \sum_{\alpha\beta} \langle e^{-\ibq \cdot (\bm{r}_\alpha - \bm{r}_\beta)} \rangle
    = -\sum_{\alpha \beta} \big\langle \bm{v}_\alpha(\Delta t) e^{-\ibq \cdot (\bm{r}_\alpha(\Delta t) - \bm{r}_\beta(0))} \big\rangle \,.
\end{equation}
Recalling that $\mathbf{D}^\mathrm{c}$ is real, Eq.~\eqref{eq-SI:flux-hypothesis-diffusion} may be reexpressed as
\begin{align}\label{eq-SI:imaginary-correlator-2}
    \begin{split}
        \mathbf{D}^\mathrm{c} \cdot \bq  &= - \Im \big[ \langle \hat{\rho}^{\bq} \hat{\rho}^{-\bq} \rangle^{-1} \langle \hat{\bm{J}}^{\bq}(\Delta t) \hat{\rho}^{-\bq}(0) \rangle \big] \\
        &= \frac{\sum_{\alpha\beta} \langle \bm{v}_\alpha(\Delta t) \sin \big( \bq \cdot [\bm{r}_\alpha(\Delta t) - \bm{r}_\beta(0)] \big) \rangle}
        {\sum_{\alpha\beta} \langle \cos\big(\bq \cdot [\bm{r}_\alpha - \bm{r}_\beta] \big)\rangle }\,,
    \end{split}
\end{align}
where $\Im[\bm{y}]$ denotes the imaginary part of $\bm{y} \in \mathbb{C}^d$.
Here we have used the fact that the density-density correlation $\langle \rho^{\bq} \rho^{-\bq} \rangle$ is real, as all particles are indistinguishable and thus the system is identical under exchange of particles $\alpha$ and $\beta$.
With these arguments, the Green-Kubo relation in Eq.~\eqref{eq-SI:collective-diffusion-green-kubo} leads to
\begin{equation}\label{eq-SI:collective-diffusion-green-kubo2}
    \mathbf{D}^\mathrm{c} = \frac{\sum_{\alpha\beta} \int_0^{\Delta t} dt\ \langle \bm{v}_\alpha(t) \bm{v}_\beta(0) \cos\big(\bm{q} \cdot [ \bm{r}_\alpha(t) - \bm{r}_\beta(0)]\big)\rangle}
    {\sum_{\alpha\beta} \langle \cos\big(\bq \cdot [\bm{r}_\alpha - \bm{r}_\beta] \big)\rangle }\,.
\end{equation}

In the dilute limit, the terms in the sums for which $\alpha \ne \beta$ vanish, and Eq.~\eqref{eq-SI:imaginary-correlator-1} reduces to
\begin{equation}\label{eq-SI:imaginary-correlator-dilute-1}
    \mathbf{D}^\mathrm{c} \cdot \ibq
    \stackrel{\text{dilute}}{=} -\langle \bm{v}(\Delta t) e^{-\ibq \cdot \bm{\Delta r} (\Delta t) } \rangle\,,
\end{equation}
or, taking the imaginary part of both sides,
\begin{equation}\label{eq-SI:imaginary-correlator-dilute-2}
    \mathbf{D}^\mathrm{c} \cdot \bq
    \stackrel{\text{dilute}}{=} \langle \bm{v}(\Delta t) \sin \big(\bm{q} \cdot \bm{\Delta r} (\Delta t) \big) \rangle\,.
\end{equation}
Here, $\bm{\Delta r}_\alpha (\Delta t) = \bm{r}_\alpha (\Delta t) - \bm{r}_\alpha(0)$, and where we have assumed for simplicity that all particles are identical, thus removing the summations.
The corresponding Green-Kubo relations, obtained by differentiating and then integrating in time, are
\begin{equation}
    \mathbf{D}^\mathrm{c}(\bq, \Delta t)
    \stackrel{\text{dilute}}{=} \int_0^{\Delta t} dt\ \bigg\langle \bm{v}(t) \otimes \bm{v}(0) \cos \big(\bm{q} \cdot \bm{\Delta r} (t) \big) \bigg\rangle\,,
\end{equation}

Using the self-similar form of Eq.~\eqref{eq:Green-Kubo-limit}, we find the macroscopic collective diffusion to be
\begin{align}\label{eq-SI:dilutediff}
    \begin{split}
        \mathbf{D}^\mathrm{c} &\stackrel{\text{dilute}}{=} \lim_{\Delta t \rightarrow 0} \lim_{\lambda \rightarrow 0}
        \int_0^{\Delta t / \lambda^2} dt\ \bigg\langle \bm{v}(t) \bm{v}(0) \cos \big(\lambda \bm{q} \cdot \bm{\Delta r} (t) \big) \bigg\rangle \\
        &= \lim_{\Delta t \rightarrow 0} \lim_{\lambda \rightarrow 0}
        \int_0^{\Delta t / \lambda^2} dt\ \bigg\langle \bm{v}(t) \bm{v}(0) \big[1 - \frac{1}{2}\lambda^2 \bm{q} \cdot \bm{\Delta r}(t) \bm{\Delta r}(t) \cdot \bm{q} + \mathcal{O}(\lambda^4)\big] \bigg\rangle \\
        = \int_0^\infty dt\ \langle \bm{v}(t) \bm{v}(0)\rangle
        &- \frac{1}{2}\lim_{\Delta t \rightarrow 0} \lim_{\lambda \rightarrow 0}
        \bigg[
        \lambda^2 \int_0^{\tau_c} dt\ \bigg\langle \bm{v}(t) \bm{v}(0)   \bm{q} \cdot \bm{\Delta r}(t) \bm{\Delta r}(t) \cdot \bm{q} \bigg\rangle\\
        &\hspace{1.8cm}+  \lambda^2  \int_{\tau_c}^{\Delta t / \lambda^2} dt\ \big\langle \bm{v}(t) \bm{v}(0) \big\rangle   \bm{q} \cdot \big\langle \bm{\Delta r}(t) \bm{\Delta r}(t) \big\rangle \cdot \bm{q} \bigg]\\
        &\hspace{-3cm}= \int_0^\infty dt\ \langle \bm{v}(t) \bm{v}(0)\rangle\,,
    \end{split}
\end{align}
recovering the Green-Kubo relation for self-diffusion derived in our previous work~\cite{Hargus2021}.
In deriving Eq.~\eqref{eq-SI:dilutediff}, on the second line, the cosine term has been expanded in $\lambda$.
On the following line, the zeroth order term in the expansion yields the expected Green-Kubo relation, with the upper limit of integration going to $\infty$ upon applying the limit $\lambda \rightarrow 0$.
The remaining integral is then separated into an integral for $t \le \tau_c$, and another for $t > \tau_c$.
The former integral is finite, and thus vanishes under the limit $\lambda \rightarrow 0$.
In the latter, where $t > \tau_c$, the velocities become uncorrelated with the squared displacement.
With the assumption that $\langle \bm{v}(t) \bm{v}(0) \rangle = 0$ for $t > \tau_c$, this integral also vanishes.
More precisely, because the mean squared displacement $\langle \bm{\Delta r}(t) \bm{\Delta r}(t) \rangle$ grows linearly with $t$, the velocity correlation $\langle \bm{v}(t) \bm{v}(0) \rangle = 0$ must decay faster than $1/t$ as $t \rightarrow \infty$, which is already the criterion for integrability of the velocity autocorrelation function.

\subsection{Self-Diffusion}\label{section:appendix-odd-diffusion-dilute-PRL}
Macroscopic transport phenomena typically arise out of complex, many-body microscopic dynamics, which generally do not admit an exact solution, but can instead be understood in terms of statistical averages and fluctuations.
In this way, the flux hypothesis provides a necessary connection between microscopic dynamics and macroscopic constitutive phenomena, providing a starting place for deriving Green-Kubo relations, just as Onsager's regression hypothesis has done for the subset of transport phenomena excluding odd transport.

Dilute self-diffusion, however, can be understood in terms of the stochastic dynamics of a single solute particle.
Accordingly, as previously shown in~\cite{Hargus2021}, Green-Kubo relations for dilute self-diffusivity (including odd diffusivity) can be derived without the flux hypothesis, by assuming only separation of timescales, as in Eq.~\eqref{eq:separation-of-timescales}.
In this appendix, we provide a self-contained derivation of the Green-Kubo relation for the dilute odd diffusivity, following Ref.~\cite{Hargus2021}, which does not require invoking the flux hypothesis.
Our motivation is to highlight parallels between this derivation and the more general one which uses the flux hypothesis, as demonstrated in the following appendix for odd collective diffusion.

To set the stage, we first show the canonical derivation of the mean-squared displacement relation for the diffusivity, which neglects the odd part.
This canonical derivation considers the time evolution of the probability density $P(\bm{r}, t | \bm{r}_0)$ of the particle being at position $\bm{r}$ at time $t$ conditioned on having started at $\bm{r}_0$ at time $0$.
This probability is then expanded by considering all sources of incoming probability from a distance $\bm{\Delta r}$ away at a time $\Delta t$ in the past
\begin{equation}\label{eq-SI:einstein}
    P(\bm{r}, t+\Delta t| \bm{r}_0) = \int d\bm{\Delta r} P(\bm{r} - \bm{\Delta r}, t| \bm{r}_0) \Pi(\bm{\Delta r}; \Delta t)\,,
\end{equation}
where $\Pi(\bm{\Delta r}; \Delta t)$ is the transition probability, i.e. the probability of the particle undergoing a displacement $\bm{\Delta r}$ in the time interval $\Delta t$.
The Markovian form of Eq.~\eqref{eq-SI:einstein} requires $\Delta t \gg \tau_c$ where $\tau_c$ is the timescale on which increments of the particle motion are correlated.
Assuming now that $\Delta t$ also satisfies $\Delta t \ll \tau_r$, as in Eq.~\eqref{eq:separation-of-timescales}, we can expand the LHS of Eq.~\eqref{eq-SI:einstein} to first order in time and the RHS to second order in space, yielding
\begin{align}
    \begin{split}
        &P(\bm{r}, t| \bm{r}_0) + \Delta t \partial_t P(\bm{r}, t| \bm{r}_0)\\
        &= \int d\bm{\Delta r} \left[ P(\bm{r}, t| \bm{r}_0) - \bm{\Delta r} \cdot \bm{\nabla} P(\bm{r}, t| \bm{r}_0) + \frac{1}{2} \bm{\Delta r} \cdot \bm{\nabla} \bm{\nabla} P(\bm{r}, t| \bm{r}_0) \cdot \bm{\Delta r} \right] \Pi(\bm{\Delta r}; \Delta t)
        + \mathcal{O}(\bm{\Delta r}^3, \Delta t^2)\\
        &= P(\bm{r}, t| \bm{r}_0) - \langle \bm{\Delta r}(\Delta t) \rangle \cdot \bm{\nabla} P(\bm{r}, t| \bm{r}_0) + \frac{1}{2} \langle \bm{\Delta r}(\Delta t) \bm{\Delta r}(\Delta t) \rangle : \bm{\nabla} \bm{\nabla} P(\bm{r}, t| \bm{r}_0)
        + \mathcal{O}(\bm{\Delta r}^3, \Delta t^2)\,.
    \end{split}
\end{align}
Dividing both sides by $\Delta t$ and neglecting the higher order terms then yields the relaxation equation
\begin{equation}
    \partial_t P(\bm{r}, t| \bm{r}_0) = - \bar{\bm{v}} \cdot \bm{\nabla} P(\bm{r}, t| \bm{r}_0) + \bm{\nabla} \cdot \mathbf{D}^\mathrm{self,\, even} \cdot \bm{\nabla} P(\bm{r}, t| \bm{r}_0)\,,
\end{equation}
where $\bar{\bm{v}} = \langle \bm{\Delta r}(\Delta t) \rangle / \Delta t$ is the drift velocity and the symmetric (i.e. even) part of the diffusivity tensor obeys
\begin{align}\label{eq-SI:GK-self-diffusivity-even}
    \begin{split}
        \mathbf{D}^\mathrm{self,\, even}
         &= \lim_{\Delta t \rightarrow \infty} \frac{1}{2 \Delta t} \langle \bm{\Delta r}(\Delta t) \otimes \bm{\Delta r}(\Delta t) \rangle^\mathrm{even}\\
         &= \int_0^\infty dt\ \langle \bm{v}(t) \otimes \bm{v}(0) \rangle^\mathrm{even}\,,
    \end{split}
\end{align}
upon sending $\Delta t \rightarrow \infty$.
The first equality is the well-known mean-squared displacement formula~\cite{VanKampen1981}.
In the final equality, defining $\bm{\Delta r} = \int_0^{\Delta t} dt\ \bm{v}(t)$ obtains the corresponding Green-Kubo formula for the even part of the diffusivity.

The time evolution of $P(\bm{r}, t| \bm{r}_0)$, however, is agnostic to odd diffusion.
To derive the Green-Kubo relation for the odd self-diffusivity, we must instead consider the fluxes---i.e. the constitutive behavior---directly, as in~\cite{Hargus2021}.
We now consider the joint probability density $P(\bm{r}, \bm{v}, t| \bm{r}_0)$, of the particle being at $\bm{r}$ with velocity $\bm{v}$ at time $t$, conditioned on having been at $\bm{r}_0$ at time $0$.
The evolution of the density $\rho(\bm{r}, t| \bm{r}_0)$ and the flux $\bm{J}(\bm{r}, t| \bm{r}_0)$, given that the particle started at $\bm{r}_0$, are then the zeroth and first moments of the joint probability density with respect to the velocity:
\begin{align}
    \label{eq-SI:zeroth-moment}
    \rho(\bm{r}, t | \bm{r}_0) &= \int d\bm{v}\ P(\bm{r}, \bm{v}, t| \bm{r}_0)\,,\\ 
    \label{eq-SI:first-moment}
    \bm{J}(\bm{r}, t | \bm{r}_0) &= \int d\bm{v}\ \bm{v} P(\bm{r}, \bm{v}, t| \bm{r}_0)\,.
\end{align}
We now introduce into the joint probability the variable $\bm{r}'$ representing the position of the particle at time $t-\Delta t$, and marginalize it back out;
that is, ${P(\bm{r}, \bm{v}, t| \bm{r}_0) = \int d\bm{r}' \ P(\bm{r}, \bm{v}, t; \bm{r}', t-\Delta t | \bm{r}_0)}$.
The expression for the flux is then
\begin{align}
    \begin{split}
        \bm{J}(\bm{r}, t| \bm{r}_0) &= \iint d\bm{v}\ d\bm{r}'\ \bm{v} P(\bm{r}, \bm{v}, t; \bm{r}', t-\Delta t | \bm{r}_0) \\
        &= \iint d\bm{v}\ d\bm{r}'\ \bm{v}
        \rho(\bm{r}', {t-\Delta t}| \bm{r}_0) P(\bm{r}, \bm{v}, t | \bm{r}', t-\Delta t; \bm{r}_0)\,,
    \end{split}
\end{align}
where the second equality uses the factorization property of the conditional probability and the third replaces $P(\bm{r}', t-\Delta t | \bm{r}_0)$ with $\rho(\bm{r}', {t-\Delta t}| \bm{r}_0)$ from Eq.~\eqref{eq-SI:zeroth-moment}.
So far, no assumptions have been made on the dynamics of the particle.
We now require $\Delta t \gg \tau_c$ (where $\tau_c$ is the timescale on which the random motion is correlated with itself) such that the dynamics are Markovian.
That is,
\begin{equation}
P(\bm{r}, \bm{v}, t | \bm{r}',  t-\Delta t; \bm{r}_0) =
P(\bm{r}, \bm{v}, t | \bm{r}',  t-\Delta t) =:
\Pi(\bm{\Delta r}, \bm{v}; \Delta t)
\end{equation}
is the transition probability of the particle undergoing a displacement $\bm{\Delta r} = \bm{r} - \bm{r}'$ and ending with velocity $\bm{v}$ in a time interval $\Delta t$, and is independent of $\bm{r}_0$.
Making the change of variables $\bm{\Delta r} = \bm{r} - \bm{r}'$, this leaves
\begin{align}
    \begin{split}
        \bm{J}(\bm{r}, t| \bm{r}_0) &= \iint d\bm{v}\ d\bm{\Delta r} \ \bm{v}
   \rho(\bm{r} - \bm{\Delta r}, t - \Delta t| \bm{r}_0)
   \Pi(\bm{\Delta r}, \bm{v}; \Delta t)
   \\
   &= \iint d\bm{v}\ d\bm{\Delta r} \ \bm{v} \big[ \rho(\bm{r}, t| \bm{r}_0) - \bm{\Delta r} \cdot \bm{\nabla} \rho(\bm{r},t| \bm{r}_0) \big] \Pi(\bm{\Delta r}, \bm{v}; \Delta t)
   + \mathcal{O}(\bm{\Delta r}^2, \Delta t)
   \\
   &= \langle \bm{v} \rangle \rho(\bm{r}, t| \bm{r}_0) - \langle \bm{v} \otimes \bm{\Delta r} \rangle \cdot \bm{\nabla} \rho(\bm{r}, t| \bm{r}_0)
   + \mathcal{O}(\bm{\Delta r}^2, \Delta t)\\
   &= \bar{\bm{v}} \rho(\bm{r}, t| \bm{r}_0) - \mathbf{D}^\mathrm{self} \cdot \bm{\nabla} \rho(\bm{r},t| \bm{r}_0)
   + \mathcal{O}(\bm{\Delta r}^2, \Delta t)\,.
    \end{split}
\end{align}
In the second equality, $\rho(\bm{r} - \bm{\Delta r}, t - \Delta t| \bm{r}_0)$ has been expanded to first order in the position and zeroth order in time, under the assumption that $\Delta t \ll \tau_r$, thus requiring the separation of timescales as in Eq.~\eqref{eq:separation-of-timescales}. The final line reveals a drift term $\bar{\bm{v}} = \langle \bm{v} \rangle$ and the diffusivity
\begin{align}\label{eq-SI:self-diffusion-GK}
    \begin{split}
    \mathbf{D}^\mathrm{self} &= \lim_{t \rightarrow \infty} \langle \bm{v}(t) \otimes \bm{\Delta r}(t) \rangle\\
    &= \lim_{t \rightarrow \infty} \int_0^t dt'\ \langle \bm{v}(t) \otimes \bm{v}(t') \rangle\\
    &= \lim_{t \rightarrow \infty} \int_0^t dt'\ \langle \bm{v}(t-t') \otimes \bm{v}(0) \rangle\\
    &= \int_0^\infty dt\ \langle \bm{v}(t) \otimes \bm{v}(0) \rangle\,.
    \end{split}
\end{align}
The final equality gives the desired Green-Kubo relation, which follows from invoking stationarity in the third equality and performing a change of variables in the final equality.
Note that Eq.~\eqref{eq-SI:self-diffusion-GK} holds for both the antisymmetric (odd) as well as the symmetric (even) part of $\mathbf{D}$.
Considering a two-dimensional isotropic system where the diffusivity takes the form $\mathbf{D} = D_\parallel \bm{\delta} - D_\perp \bm{\epsilon}$,
we arrive at the Green-Kubo relations
\begin{align}
    \label{eq-SI:self-diffusion-GK-even}
    D^{\mathrm{self}}_\parallel &= \frac{1}{2} \int_0^\infty dt\ \langle \bm{v}(t) \cdot \bm{v}(0) \rangle
    = \lim_{t \rightarrow \infty} \frac{1}{4t}\langle |\bm{\Delta r}(t)|^2 \rangle\,, \\
    \label{eq-SI:self-diffusion-GK-odd}
    D^{\mathrm{self}}_\perp &= -\frac{1}{2} \int_0^\infty dt\ \langle \bm{v}(t) \bm{v}(0) \rangle : \bm{\epsilon}\\
    &= \frac{1}{2} \int_0^\infty dt\ \langle \bm{v}_y(t) \bm{v}_x(0) \rangle - \langle \bm{v}_x(t) \bm{v}_y(0) \rangle\,,\notag
\end{align}
Equation~\eqref{eq-SI:self-diffusion-GK-odd} reveals the odd diffusivity to be nonzero only for dynamics breaking time-reversal and parity symmetries, i.e. chiral random motion.
Finally, note that while $D^{\mathrm{self}}_\parallel$ can be expressed in terms of the mean squared displacement, as in Eq.~\eqref{eq-SI:GK-self-diffusivity-even}, no such relationship exists for the odd diffusivity $D^{\mathrm{self}}_\perp$ due to its inherent antisymmetry.

\section{Dilute Inertial Langevin Dynamics}\label{section:appendix-langevin}
\subsection{One Dimension}
In this appendix, we shall derive the exact form of the correlation function in Eq.~\eqref{eq-SI:imaginary-correlator-dilute-1} for dilute Langevin diffusion. We then show that carrying out the limits in Eq.~\eqref{eq:Green-Kubo-limit} yields the standard Green-Kubo expression for self-diffusion.
Before proceeding to the two-dimensional case where odd collective diffusion is possible, we briefly solve for the correlation function in Eq.~\eqref{eq-SI:imaginary-correlator-dilute-1} in a one-dimensional system of particles subject to the underdamped Langevin dynamics
\begin{align}
    \label{eq-SI:Langevin-1d-A}
    \dot{x}_\alpha(t) &= v_\alpha(t)\,,\\
    \label{eq-SI:Langevin-1d-B}
    \dot{v}_\alpha(t) &= -\gamma v_\alpha(t) + \eta_\alpha(t)\,,
\end{align}
where $\gamma$ is the friction coefficient and $\eta_\alpha(t)$ represents Gaussian white noise satisfying $\langle \eta_\alpha(t) \rangle = 0$ and $\langle \eta_\alpha(t) \eta_\beta(t') \rangle = 2 \gamma k_\mathrm{B} T \delta(t -t')\delta_{\alpha\beta}$~\cite{Chandrasekhar1943,VanKampen1981}.

Solving Eqs.~\eqref{eq-SI:Langevin-1d-A} and~\eqref{eq-SI:Langevin-1d-B} for the particle displacement yields
\begin{align}\label{eq-SI:displacement-solution-langevin-1d}
    \begin{split}
    \Delta x(\Delta t) &= \int_0^{\Delta t} dt\ v(t) \\
    &= \int_0^{\Delta t} dt\ \bigg [e^{-\gamma t} v(0) + \int_0^{t} dt'\ e^{-\gamma (t - t')} \eta(t') \bigg] \\
    &= \big(1 - e^{-\gamma \Delta t} \big)\frac{v(0)}{\gamma} + \int_0^{\Delta t} dt\ \frac{1}{\gamma} \big(1 - e^{-\gamma(\Delta t-t)} \big) \eta(t)\,.
    \end{split}
\end{align}
where for clarity we have removed particle index $\alpha$, as the particles are identical and non-interacting.
The final equality above is obtained by changing the order of integration.
Now, using $\langle \eta(t) v(0) \rangle=0$ and the equilibrium equipartition identity $\langle |v(0)|^2 \rangle = k_\mathrm{B} T$, the mean squared displacement is
\begin{align}\label{eq-SI:msd-langevin-1d}
    \begin{split}
    \langle |\Delta x(\Delta t)|^2 \rangle &= \frac{k_\mathrm{B} T}{\gamma^2}(1 - e^{-\gamma \Delta t})^2
    + \int_0^{\Delta t} dt \int_0^{\Delta t} dt'\ \frac{1}{\gamma^2}(1 - e^{-\gamma(\Delta t - t)})(1 - e^{-\gamma(\Delta t - t')}) \langle \eta(t) \eta(t') \rangle\\
    &= \frac{k_\mathrm{B} T}{\gamma^2}(1 - e^{-\gamma \Delta t})^2 + \frac{2 k_\mathrm{B} T}{\gamma} \int_0^{\Delta t} dt\ (1 - e^{-\gamma(\Delta t - t)})^2\\
    &= \frac{2 k_\mathrm{B} T}{\gamma^2} \big[ \gamma \Delta t + e^{-\gamma \Delta t} - 1 \big]\,,
    \end{split}
\end{align}
where the self-diffusivity is given by
\begin{equation}\label{eq-SI:self-diffusion}
D^{\mathrm{self}} := \lim_{\Delta t \rightarrow \infty} \frac{1}{2 \Delta t} \langle |\Delta x(\Delta t)|^2 \rangle = \frac{k_\mathrm{B} T}{\gamma}\,.
\end{equation}

The correlation function in Eq.~\eqref{eq-SI:imaginary-correlator-dilute-1} can be expressed as the time derivative of the moment generating function
\begin{equation}\label{eq-SI:characteristic-function-1}
    D^\mathrm{c}(q,\Delta t) = -\frac{1}{\text{i} q} \langle v(\Delta t) e^{-\text{i} q \Delta x(\Delta t)} \rangle
    = -\frac{1}{q^2} \frac{\partial}{\partial \Delta t} \langle e^{- \text{i} q \Delta x(\Delta t)} \rangle \,.
\end{equation}
A cumulant expansion yields
\begin{equation}\label{eq-SI:cumulant-expansion-1d}
    \langle e^{-\text{i} q \Delta x(\Delta t)} \rangle
    = \exp \bigg\{-\frac{1}{2} q^2 \big\langle |\Delta x(\Delta t)|^2 \big \rangle \bigg\}\,,
\end{equation}
where only the second cumulant remains due to the Gaussian nature of the noise $\eta(t)$.
Putting together Eqs.~\eqref{eq-SI:msd-langevin-1d}, \eqref{eq-SI:characteristic-function-1} and~\eqref{eq-SI:cumulant-expansion-1d} results in
\begin{align}
    \begin{split}
         D^\mathrm{c}(q,\Delta t) &= - \frac{1}{q^2} \frac{\partial}{\partial \Delta t} \exp\bigg\{ -\frac{1}{2} q^2 \langle | \Delta x (\Delta t)|^2 \rangle \bigg\}\\
         &= \frac{k_\mathrm{B} T}{\gamma} [1 - e^{-\gamma \Delta t}] \exp\bigg\{ -q^2 \frac{k_\mathrm{B} T}{\gamma^2} \big[ \gamma \Delta t + e^{-\gamma \Delta t} - 1 \big] \bigg\}\,.
    \end{split}
\end{align}
Rescaling by $\lambda$ and evaluating the limit $\lambda \rightarrow 0$ recovers Markovian constitutive behavior
\begin{align}
    \begin{split}
    \lim_{\lambda \rightarrow 0} D^\mathrm{c}(\lambda q,\Delta t / \lambda^2)
    &= \lim_{\lambda \rightarrow 0} \frac{k_\mathrm{B} T}{\gamma} [1 - e^{-\gamma \Delta t/\lambda^2}] \exp\bigg\{ -\lambda^2 q^2 \frac{k_\mathrm{B} T}{\gamma^2} \big[ \gamma \Delta t/\lambda^2 + e^{-\gamma \Delta t/\lambda^2} - 1 \big] \bigg\}\\
    &= \frac{k_\mathrm{B} T}{\gamma} \exp\bigg\{ -  \frac{k_\mathrm{B} T}{\gamma} q^2 \Delta t \bigg\}\,.
    \end{split}
\end{align}
Evaluating the limit $\Delta t \rightarrow 0$ then yields
\begin{align}
    \begin{split}
        D^\mathrm{c} = \lim_{\Delta t \rightarrow 0} \lim_{\lambda \rightarrow 0} D^\mathrm{c}(\lambda q,\Delta t / \lambda^2)
        = \lim_{\Delta t \rightarrow 0} \frac{k_\mathrm{B} T}{\gamma} \exp\bigg\{ -  \frac{k_\mathrm{B} T}{\gamma} q^2 \Delta t \bigg\}
        = \frac{k_\mathrm{B} T}{\gamma}\,,
    \end{split}
\end{align}
which agrees with the self-diffusion result in Eq.~\eqref{eq-SI:self-diffusion} due to the non-interacting setting.

\subsection{Two Dimensions with Odd Friction}
We now proceed to the two dimensional case, where additional attention is needed to obtain the odd (antisymmetric) part of the diffusivity.
In two dimensions, the Eqs.~\eqref{eq-SI:Langevin-1d-A} and~\eqref{eq-SI:Langevin-1d-B} for inertial Langevin dynamics generalize to
\begin{align}
    \label{eq-SI:Langevin-2d-A}
    \dot{\bm{r}}_\alpha(t) &= \bm{v}_\alpha(t) \,, \\
    \label{eq-SI:Langevin-2d-B}
    \dot{\bm{v}}_\alpha(t) &= -\bm{\gamma} \cdot \bm{v}_\alpha(t) + \bm{\eta}_\alpha(t) \,,
\end{align}
where $\bm{\gamma} = \gamma_\parallel \bm{\delta} + \gamma_\perp \bm{\epsilon}$ is the isotropic friction on the particle due to interactions with its environment.
The odd friction $\gamma_\perp$ causes the particle motion to break time-reversal and parity symmetries.
In practice, this term could be the consequence of hydrodynamic asymmetry causing the particle to deflect from its current direction with a systematic bias~\cite{Kummel2013,Nourhani2016},
or the consequence of a charged particle moving through the plane normal to a magnetic field.
We assume the noise $\bm{\eta}_\alpha$ is related to the friction through a fluctuation dissipation relation $\langle \bm{\eta}_\alpha(t) \bm{\eta}_\beta(t') \rangle = 2 k_\mathrm{B} T \bm{\gamma} \delta_{\alpha\beta} \delta(t-t')$.

Solving for the particle displacement from Eqs.~\eqref{eq-SI:Langevin-2d-A} and~\eqref{eq-SI:Langevin-2d-B} yields
\begin{align}\label{eq-SI:displacement-solution-langevin-2d}
    \begin{split}
    \bm{\Delta r}(\Delta t) &= \int_0^{\Delta t} dt\ \bm{v}(t) \\
    &= \int_0^{\Delta t} dt\ \bigg [e^{-\bm{\gamma} t} \cdot \bm{v}(0) + \int_0^{t} dt'\ e^{-\bm{\gamma} (t - t')} \cdot \bm{\eta}(t') \bigg] \\
    &= \bm{\gamma}^{-1} \big(\bm{\delta} - e^{-\bm{\gamma} \Delta t} \big) \cdot \bm{v}(0) + \bm{\gamma}^{-1}  \int_0^{\Delta t} dt\ \big(\bm{\delta} - e^{-\bm{\gamma}(\Delta t-t)} \big) \cdot \bm{\eta}(t) \\
    &= \mathbf{G}(\Delta t) \cdot \bm{v}(0) + \int_0^{\Delta t} dt\ \mathbf{G}(\Delta t - t) \cdot \bm{\eta}(t)\,,
    \end{split}
\end{align}
where the particle index $\alpha$ has been ignored due to particles being identical and non-interacting.
The final line defines the Green's function 
$\mathbf{G}(\Delta t) = \bm{\gamma}^{-1} \big(\bm{\delta} - e^{-\bm{\gamma} \Delta t} \big)$.
It then follows that the first and second moments of the displacement, conditioned on an initial particle velocity $\bm{v}(0)$, are
\begin{align}
    \label{eq-SI:conditional-moment-1}
    \langle \bm{\Delta r} (\Delta t)\rangle_{\bm{v}(0)} &= \mathbf{G}(\Delta t) \cdot \bm{v}(0) \,, \\
    \label{eq-SI:conditional-moment-2}
    \langle \bm{\Delta r}(\Delta t) \bm{\Delta r}(\Delta t) \rangle_{\bm{v}(0)}
    &= \mathbf{G}(\Delta t) \big( \bm{v}(0) \bm{v}(0) \big) \mathbf{G}^\intercal(\Delta t) \notag \\
    &+ 2 k_\mathrm{B} T {\bm{\gamma}^\intercal}^{-1} \bigg[ \Delta t  - \Tr \mathbf{G}(\Delta t) + \frac{1}{2\gamma_\parallel}\big( 1 - e^{-2\gamma_\parallel\Delta t}\big)\bigg] \,.
\end{align}
The initial velocity $\bm{v}(0)$ is distributed in the steady state corresponding to Eqs.~\eqref{eq-SI:Langevin-2d-A} and~\eqref{eq-SI:Langevin-2d-B} by the Maxwell-Boltzmann distribution
\begin{equation}\label{eq-SI:maxwell-boltzmann}
    W_{ss}(\bm{v}_0) = \frac{1}{2 \pi k_\mathrm{B} T} \exp\left[ -\frac{|\bm{v}_0|^2}{2 k_\mathrm{B} T} \right] \,.
\end{equation}

From isotropy of the dynamics, the diffusivity tensor will be of the form $\mathbf{D}^\mathrm{c} = D^\mathrm{c}_\parallel \bm{\delta} - D^\mathrm{c}_\perp \bm{\epsilon}$, where the presence of the odd diffusivity $D^\mathrm{c}_\perp$ renders the situation more complicated than the one-dimensional case solved in the previous section.
Under time-reversal of the dynamics, $D^\mathrm{c}_\parallel$ remains unchanged while $D^\mathrm{c}_\perp$ changes sign, as can be seen from the Green-Kubo relation~\eqref{eq:GK-collective-diffusion}.
Thus, we shall consider the time-reversed diffusivity ${\mathbf{D}^\mathrm{c}}^\intercal = D^\mathrm{c}_\parallel \bm{\delta} + D^\mathrm{c}_\perp \bm{\epsilon}$, taking advantage of the fact that it can be readily computed from the time-reversed form of Eq.~\eqref{eq-SI:imaginary-correlator-dilute-1}, i.e., 
\begin{align}\label{eq-SI:correlation-transition-2}
    \begin{split}
        &\mathbf{D}^\mathrm{c}(\bq, \Delta t) \cdot \ibq = - \langle \bm{v}(\Delta t) e^{-\ibq \cdot \bm{\Delta r}(\Delta t))} \rangle\\
        \xRightarrow{\mathcal{T}} &{\mathbf{D}^\mathrm{c}}^\intercal(\bq, \Delta t) \cdot \ibq
        = \langle \bm{v}(0) e^{\ibq \cdot \bm{\Delta r}(\Delta t))} \rangle
        = \int d\bm{v}_0\ W_{ss}(\bm{v}_0) \bm{v}_0 \langle e^{\ibq \cdot \bm{\Delta r}(\Delta t))} \rangle_{\bm{v}_0} \,,
    \end{split}
\end{align}
where $\xRightarrow{\mathcal{T}}$ indicates the time-reversal operation and $\langle \cdot \rangle_{\bm{v}_0}$ is the expectation conditioned on the initial velocity $\bm{v}(0) = \bm{v}_0$.
It is important to note that, while in the one-dimensional case described in the previous section it was possible to obtain the $D^\mathrm{c}$ simply by time differentiation of the moment generating function, in the two-dimensional case such an approach would yield an expression only for the even part of $\mathbf{D}^\mathrm{c}$ and not the odd part. Nevertheless, one may expand the conditional moment generating function to obtain
\begin{equation}\label{eq-SI:cumulant-expansion-2d}
    \langle e^{\ibq \cdot \bm{\Delta r}(\Delta t)} \rangle_{\bm{v}_0}
    =\exp \bigg\{ \ibq \cdot \langle \bm{\Delta r}(\Delta t) \rangle_{\bm{v}_0}\bigg\}
    \exp \bigg\{-\frac{1}{2}\bq \cdot \big\langle \big( \bm{\Delta r}(\Delta t) \bm{\Delta r}(\Delta t) - \langle \bm{\Delta r}(\Delta t)\rangle_{\bm{v}_0} \langle \bm{\Delta r}(\Delta t)\rangle_{\bm{v}_0} \big) \big\rangle_{\bm{v}_0} \cdot \bq \bigg\}\,.
\end{equation}
As in the one-dimensional case, all higher cumulants vanish due to the Gaussian nature of the noise. Due to the initial velocity condition $\bm{v}(0) = \bm{v}_0$, however, the first cumulant $\langle \bm{\Delta r}(\Delta t) \rangle_{\bm{v}_0}$ does not vanish.
Using the identities in Eqs.~\eqref{eq-SI:conditional-moment-1}-\eqref{eq-SI:conditional-moment-2}, the second cumulant simplifies as
\begin{align}\label{eq-SI:2nd-cumulant}
    \begin{split}
        \bq \cdot \bigg\langle \big( \bm{\Delta r}(\Delta t) \bm{\Delta r}(\Delta t) - \langle \bm{\Delta r}(\Delta t)\rangle_{\bm{v}_0} \langle \bm{\Delta r}(\Delta t)\rangle_{\bm{v}_0} \big) \bigg\rangle_{\bm{v}_0} \cdot \bq
        &= \frac{1}{2}q^2 \big\langle | \bm{\Delta r}(\Delta t)|^2 \big\rangle
        - \bq \cdot \mathbf{G}(\Delta t) \langle \bm{v}_0 \bm{v}_0 \rangle \mathbf{G}^\intercal(\Delta t) \cdot \bq\\
        &= \frac{1}{2} q^2 \big\langle | \bm{\Delta r}(\Delta t)|^2 \big\rangle
        - k_\mathrm{B} T \bq \cdot \mathbf{G}(\Delta t) \mathbf{G}^\intercal(\Delta t) \cdot \bq\,.
    \end{split}
\end{align}
In the first equality, by recognizing from Eq.~\eqref{eq-SI:conditional-moment-2} that the second cumulant does not depend on the initial velocity, we average each term individually over the initial velocity according to the distribution $W_{ss}(\bm{v}_0)$, from which $\langle \bm{v}(0) \bm{v}(0) \rangle = k_\mathrm{B} T \bm{\delta}$.
With Eq.~\eqref{eq-SI:2nd-cumulant}, Eq.~\eqref{eq-SI:correlation-transition-2} is then a Gaussian integral which evaluates to
\begin{align}\label{eq-SI:correlation-transition-3}
    \begin{split}
        {\mathbf{D}^\mathrm{c}}^\intercal(\bq, \Delta t) \cdot \ibq
        &= \int d\bm{v}_0\ W_{ss}(\bm{v}_0)\ \bm{v}_0
        \exp \bigg\{\ibq \cdot \mathbf{G}(\Delta t) \cdot \bm{v}_0\bigg\}
    \exp \bigg\{-\frac{1}{4} q^2 \big\langle | \bm{\Delta r}(\Delta t)|^2 \big\rangle + \frac{k_\mathrm{B} T}{2} \bq \cdot \mathbf{G}(\Delta t) \mathbf{G}^\intercal(\Delta t) \cdot \bq \bigg\} \\
        &= e^{-\frac{1}{4}q^2 \langle | \bm{\Delta r}(\Delta t)|^2 \rangle}
        \frac{1}{2\pi k_\mathrm{B} T} \int d\bm{v}_0\ \bm{v}_0 \exp\bigg\{ -\frac{1}{2 k_\mathrm{B} T} |\bm{v}_0|^2 + \ibq \cdot \mathbf{G}(\Delta t) \cdot \bm{v}_0 + \frac{k_\mathrm{B} T}{2} \bq \cdot \mathbf{G}(\Delta t) \mathbf{G}^\intercal(\Delta t) \cdot \bq \bigg\}
        \\
        &= e^{-\frac{1}{4}q^2 \langle | \bm{\Delta r}(\Delta t)|^2 \rangle}
        \frac{1}{2\pi k_\mathrm{B} T} \int d\bm{v}_0\ \bm{v}_0 \exp\bigg\{ -\frac{1}{2 k_\mathrm{B} T} \big|\bm{v}_0 - k_\mathrm{B} T \mathbf{G}^\intercal(\Delta t) \cdot \ibq   \big|^2 \bigg\}\\
        &=  k_\mathrm{B} T \mathbf{G}^\intercal (\Delta t) \cdot \ibq\,
        e^{-\frac{1}{4}q^2 \langle | \bm{\Delta r}(\Delta t)|^2 \rangle} \,.
    \end{split}
\end{align}
From isotropy, the orientation of $\bq$ may be chosen arbitrarily and thus
\begin{equation}\label{eq-SI:correlation-transition-4}
    \boxed{\mathbf{D}^\mathrm{c}(\bq, \Delta t) = k_\mathrm{B} T \mathbf{G}(\Delta t) e^{-\frac{1}{4}q^2 \langle | \bm{\Delta r}(\Delta t)|^2 \rangle}}\,.
\end{equation}
From Eq.~\eqref{eq-SI:conditional-moment-2} together with equipartition identity $\langle \bm{v}(0) \bm{v}(0) \rangle = k_\mathrm{B} T \bm{\delta}$, the mean-squared displacement is
\begin{equation}
    \langle | \bm{\Delta r}(\Delta t)|^2 \rangle = k_\mathrm{B} T \mathbf{G} (\Delta t) : \mathbf{G} (\Delta t) + 4 k_\mathrm{B} T  \frac{\gamma_\parallel}{\gamma_\parallel^2 + \gamma_\perp^2} \bigg[ \Delta t  - \Tr \mathbf{G}(\Delta t) + \frac{1}{2\gamma_\parallel}\big( 1 - e^{-2\gamma_\parallel\Delta t}\big)\bigg]
\end{equation}

The odd diffusivity $D^\mathrm{c}_\perp(\bq, \Delta t) = -\frac{1}{2} \bm{\epsilon} : \mathbf{D}^\mathrm{c}(\bq, \Delta t)$ is plotted from the expression in Eq.~\eqref{eq-SI:correlation-transition-4} in Fig.~\ref{fig:fig1-assembly}(d) of the main text.
Note that $\lim_{\Delta t \rightarrow 0} \mathbf{D}^\mathrm{c}(\bq, \Delta t) = \lim_{\Delta t \rightarrow \infty} \mathbf{D}^\mathrm{c}(\bq, \Delta t) = 0$ and it is thus only on intermediate timescales $\tau_c \ll \Delta t \ll \tau_r$ the collective diffusivity can be obtained.
As indicated in the main text, this can be achieved by rescaling by $\lambda$ and evaluating the limit
\begin{equation}
        \lim_{\lambda \rightarrow 0} \mathbf{D}^\mathrm{c}(\lambda \bq, \Delta t / \lambda^2)
        = \exp\big\{ -k_\mathrm{B} T \frac{\gamma_\parallel}{\gamma_\parallel^2 + \gamma_\perp^2} q^2 \Delta t \big\} k_\mathrm{B} T \bm{\gamma}^{-1}\,,
\end{equation}
recovering Markovian behavior consistent with the macroscopic phenomenology.
Then, taking the short-time limit
\begin{equation}
    \mathbf{D}^\mathrm{c} = \lim_{\Delta t \rightarrow 0} \lim_{\lambda \rightarrow 0} \mathbf{D}^\mathrm{c}(\lambda \bq, \Delta t / \lambda^2)
    = \lim_{\Delta t \rightarrow 0} \exp\big\{- k_\mathrm{B} T \frac{\gamma_\parallel}{\gamma_\parallel^2 + \gamma_\perp^2} q^2 \Delta t \big\} k_\mathrm{B} T \bm{\gamma}^{-1}
    = k_\mathrm{B} T \bm{\gamma}^{-1}\,.
\end{equation}
Contracting with $\bm{\delta}$ and $\bm{\epsilon}$ to obtain the even and odd components, respectively, results in
\begin{align} \label{eq-SI:langevin-gk-collective}
    D^\mathrm{c}_\parallel &= {k_\mathrm{B}T} \frac{\gamma_\parallel}{\gamma_\parallel^2 + \gamma_\perp^2}\,, \\
    D^\mathrm{c}_\perp &= {k_\mathrm{B}T} \frac{\gamma_\perp}{\gamma_\parallel^2 + \gamma_\perp^2}\,.
\end{align}
This can be compared to the direct solution of the self-diffusivity from Eq.~\eqref{eq:self-diffusion} or the expression in Eq.~\eqref{eq-SI:self-diffusion-GK} derived without the flux hypothesis, where
\begin{align}\label{eq-SI:langevin-gk-self}
    \begin{split}
        \mathbf{D}^\mathrm{self} &= \int_0^\infty dt\ \langle \bm{v}(t) \bm{v}(0) \rangle \\
        &= \int_0^\infty dt\ e^{-\bm{\gamma} t} \langle \bm{v}(0) \bm{v}(0) \rangle
        + \int_0^t dt'\ e^{-\bm{\gamma}(t-t')} \langle \bm{\eta}(t') \bm{v}(0) \rangle  \\
        &= k_\mathrm{B} T \bm{\gamma}^{-1}\,,
    \end{split}
\end{align}
with components
\begin{align}
    D^{\mathrm{self}}_\parallel &= {k_\mathrm{B}T} \frac{\gamma_\parallel}{\gamma_\parallel^2 + \gamma_\perp^2}\,, \\
    D^{\mathrm{self}}_\perp &= {k_\mathrm{B}T} \frac{\gamma_\perp}{\gamma_\parallel^2 + \gamma_\perp^2}\,.
\end{align}
As expected due to the dilute (i.e. non-interacting) setting, the collective diffusivity recovers the self diffusivity when evaluated correctly in the macroscopic, intermediate-timescale regime. Note that Fig.~\ref{fig:fig1-assembly}(c) plots several typical trajectories of Eqs.~\eqref{eq-SI:Langevin-2d-A}-\eqref{eq-SI:Langevin-2d-B} with $k_\mathrm{B} T = \gamma_\parallel = \gamma_\perp = 1$.
These same parameters are used in Figs.~\ref{fig:fig1-assembly}(d) and~\ref{fig:fig1-assembly}(e), where $t \in [0,12]$ and $t \in [0,4]$, respectively, and where colorbars are scaled logarithmically with $\bq$ and $\lambda \in [0.01, 1.5]$.

\section{Simulation Details}\label{section:appendix-simulation-details}
Molecular dynamics simulations were carried out using the LAMMPS simulation software~\cite{lammps} with custom modifications\footnote{Available at https://github.com/mandadapu-group/active-matter.}.
As a minimal model of the interacting active spinners described by Eq.~\eqref{eq:spinners-eom}, we simulated rigid dumbbell particles constructed from two beads of width $\sigma$ held together by a rigid bond of length $\ell$.
These beads interact with one another through a Weeks-Chandler-Andersen potential, defined by
\begin{equation}
V^\mathrm{WCA}(r) = \begin{cases}
      4\epsilon \bigg[ \big(\sigma/r\big)^{12} - \big(\sigma/r\big)^6 \bigg] + \epsilon & r < 2^{1/6}\sigma \\
      0 & r \geq 2^{1/6}\sigma \,, \\
   \end{cases}
\end{equation}
where $r = |\bm{r}_\alpha^a - \bm{r}_\beta^b|$ is the distance between bead $a$ in dumbbell $\alpha$ and bead $b$ in dumbbell $\beta$.
The position of each dumbbell is defined as its center of mass $\bm{r}_\alpha = \frac{1}{2}(\bm{r}_\alpha^1 +\bm{r}_\alpha^2)$, and the angle $\theta_\alpha$ of the dumbbell relative to the $x$-axis is defined such that
$\bm{\Delta r}_\alpha = \bm{r}_\alpha^2 - \bm{r}_\alpha^1 = \ell \begin{bmatrix} \cos \theta_\alpha \\ \sin \theta_\alpha \end{bmatrix}$.
Conversely, the bead positions are related to the new coordinates by $\bm{r}_\alpha^1 = \bm{r}_\alpha - \frac{1}{2} \bm{\Delta r}_\alpha$ and $\bm{r}_\alpha^2 = \bm{r}_\alpha + \frac{1}{2} \bm{\Delta r}_\alpha$. The force on the dumbbell center of mass is then
\begin{equation}
    -\frac{\partial V^\mathrm{WCA}}{\partial \bm{r}_\alpha} =
    -\bigg(\frac{\partial \bm{r}_\alpha^1}{\partial \bm{r}_\alpha} \frac{\partial}{\partial \bm{r}_\alpha^1} +
    \frac{\partial \bm{r}_\alpha^2}{\partial \bm{r}_\alpha} \frac{\partial }{\partial \bm{r}_\alpha^2} \bigg) V^\mathrm{WCA}
    = -\bigg(\frac{\partial}{\partial \bm{r}_\alpha^1} + \frac{\partial}{\partial \bm{r}_\alpha^2}\bigg)V^\mathrm{WCA}\,,
\end{equation}
and the torque is
\begin{align}
    \begin{split}
        -\frac{\partial V^\mathrm{WCA}}{\partial \theta_\alpha} &= -\frac{\partial \bm{\Delta r}_\alpha}{\partial \theta_\alpha} \cdot
        \bigg(\frac{\partial \bm{r}_\alpha^1}{\partial \bm{\Delta r}_\alpha} \frac{\partial V^\mathrm{WCA}}{\partial \bm{r}_\alpha^1} +
        \frac{\partial \bm{r}_\alpha^2}{\partial \bm{\Delta r}_\alpha} \frac{\partial V^\mathrm{WCA}}{\partial \bm{r}_\alpha^2}\bigg) \\
        &= \frac{\ell}{2}\bigg[ \sin \theta \bigg( \frac{\partial V^\mathrm{WCA}}{\partial x_\alpha^2} - \frac{\partial V^\mathrm{WCA}}{\partial x_\alpha^1} \bigg)
        -\cos \theta \bigg( \frac{\partial V^\mathrm{WCA}}{\partial y_\alpha^2} - \frac{\partial V^\mathrm{WCA}}{\partial y_\alpha^1} \bigg)
        \bigg]\,.
    \end{split}
\end{align}
The Green-Kubo estimates for $\mathbf{D}^\mathrm{c}$ are then computed from Eq.~\eqref{eq:GK-collective-diffusion-2} in a periodic simulation box of size $L=120\sigma$ for the wave vector $|\bm{q}| = \frac{2 \pi}{L}$.
The Green-Kubo values reported in Fig.~\ref{fig:dumbbell-results} are computed as the average over $60$ independent replicas, while 95\% confidence intervals (i.e. error bars) are computed as twice the standard error over these replicas.
All simulations use the settings $\gamma = 2,\ \gamma_\theta = \frac{1}{2},\ m = 2,\ I=\frac{1}{2},\ \ell=1,\ k_\mathrm{B}T = 1,\ \sigma=1,$ and $\epsilon=1$.
The translational and angular velocity evolution in~\eqref{eq:spinners-eom} was discretized with Euler-Maruyama method and a time step of $\Delta t = 0.005$.

As described in the main text, the Green-Kubo results are compared against direct ``non-equilibrium molecular dynamics'' (NEMD) measurements of $\mathbf{D}^\mathrm{c}$ made by perturbing the system with a small sinusoidal potential and measuring the response.
Specifically, an external potential $V^\mathrm{ext} = -\delta V \cos (qx)$ is imposed with $\delta V = 0.2 \bar{\rho}$ and $q = \frac{2 \pi}{L}$.
This potential drives a flux of the spinners relative to the substrate, independent of the flux driven by density gradients. The result is a
drift-diffusion equation for the density evolution of this single component system is then
\begin{equation}\label{eq-SI:drift-diffusion}
\bm{J} = -\mathbf{D}^\mathrm{c} \bm{\nabla} \rho + \gamma^{-1} \rho \mathbf{F}^{\mathrm{ext}}\,,
\end{equation}
where $\mathbf{F}^{\mathrm{ext}} = -\bm{\nabla} V^\mathrm{ext} = -q \delta V \sin (qx)\hat{\bm{e}}_x$.
For small $\delta V$, the density can be linearized as $\rho(\bm{r}) = \bar{\rho} + \delta \rho(\bm{r})$, such that the external force acts by $\bar{\rho}$ to leading order.
In the steady state, $\dot{\rho} = -\bm{\nabla}\cdot \bm{J} = 0$, requiring $J_x = 0$.
Solving Eq.~\eqref{eq-SI:drift-diffusion} for the steady state then yields
\begin{align}
    \rho(x) &= \bar{\rho}\bigg(1 + \frac{\delta V}{D_\parallel^\mathrm{c} \gamma} \cos(qx) \bigg)\,,\\
    J_x &= 0\,,\\
    J_y &= -D_\perp^\mathrm{c} \partial_x \rho = \frac{D_\perp^\mathrm{c} \bar{\rho} q \delta V}{D_\parallel^\mathrm{c} \gamma} \sin(q x)\,.
\end{align}
Finally, given $|\bq|=\frac{2\pi}{L}$ and the sinusoidal nature of the applied potential $V^{\text{ext}}$, the simulation box can be divided into intervals $x \in (0, L/4]$, $x \in (L/4, L/2]$, $x \in (L/2, 3L/4]$ and $x \in (3L/4, L)$ to improve sampling statistics. This approach yields
\begin{equation}
    \frac{\bar{\rho}L\delta V}{2\pi D_\parallel^\mathrm{c} \gamma}  = \frac{1}{4}\bigg[
    \int_0^{L/4} \delta\rho(x) dx\
    -\int_{L/4}^{L/2} \delta\rho(x) dx\
    -\int_{L/2}^{3L/4} \delta\rho(x) dx\
    +\int_{3L/4}^{L} \delta\rho(x) dx\
    \bigg]
\end{equation}
and
\begin{equation}
    \frac{D_\perp^\mathrm{c} \bar{\rho} qL \delta V}{2\pi D_\parallel^\mathrm{c} \gamma}  = \frac{1}{4}\bigg[
    \int_0^{L/4} J_y(x) dx\
    +\int_{L/4}^{3L/4} J_y(x) dx\
    -\int_{L/2}^{3L/4} J_y(x) dx\
    -\int_{3L/4}^{L} J_y(x) dx\
    \bigg]
\end{equation}
where $\delta\rho(x)$ is the steady state value measured in simulation after averaging in time and $y$.
The NEMD results reported in Fig.~\ref{fig:dumbbell-results} are averaged over 60 replicas, with error bars corresponding to 95\% confidence intervals smaller than the line widths and consequently omitted.

\section{Green-Kubo Relations for 2d Momentum and Energy Transport}\label{sec:SI-2dgk}
In this Appendix we show the utililty of the flux hypothesis to recover Green-Kubo relations for the transport coefficients governing two-dimensional (2d) momentum and energy transport, including odd viscous and odd thermal transport. In subsection~\ref{subsection:nospin}, we consider the viscous behaviors of systems coupling the stress tensor (momentum flux) $\bm{\sigma}$ to a velocity gradient $\bm{\nabla}\bm{v}$ via the fourth order viscosity tensor $\bm{\eta}^{(4)}$. In subsection~\ref{subsection:withspin} we consider viscous behaviors of systems with internal spin, by allowing for the additional coupling of the stress tensor to an internal spin angular momentum field $m$. In subsection~\ref{subsection:heat}, we consider energy transport of systems with odd heat conduction, coupling the heat flux vector $\bm{Q}$ to a temperature gradient $\bm{\nabla} T$ via the second order thermal conductivity tensor $\bm{\kappa}$. 

For higher order transport tensors such as the viscosity $\bm{\eta}^{(4)}$, we can apply the flux hypothesis using two approaches. In the first approach (used in subsection~\ref{subsection:nospin}), we consider the momentum in each coordinate as a separate quantity. Thus in 2d, we decompose the linear momentum into $\rho v_x$ and $\rho v_y$ and the momentum flux into $\sigma_{xi}$ and $\sigma_{yi}$, $i\in\{x,y\}$. Here, Latin indices now refer to spatial degrees of freedom and Einstein summation notation is used. The fourth order viscosity tensor $\eta_{ijkl}$ is accordingly unraveled into the four matrices $\eta_{xixj}, \eta_{yiyj}, \eta_{xiyj},$ $\eta_{yixj}$, and the result of Eq.~\eqref{eq:Green-Kubo}  may then be directly applied. However, in this unraveled state we are unable to apply tensor representation theorems\footnote{A tensor representation theorem reduces the number of transport coefficients by constructing a basis set corresponding to the material symmetry. For example, in two-dimensional isotropic systems, the fourth order viscosity tensor is reduced from 16 to 6 coefficients (see SI-I of Ref.~\cite{Epstein2020} for a representation theorem for isotropic tensors of order $n$ in $d$ dimensions).} to simplify the viscosity tensor into independent coefficients (\textit{e.g.} the shear and odd viscosities~\cite{Epstein2020}), as each coefficient may have contributions from multiple unraveled matrices. To use the representation theorems, we require reassembly of the Green-Kubo relations for the four matrices $\eta_{xixj}, \eta_{yiyj}, \eta_{xiyj},$ $\eta_{yixj}$, into a Green-Kubo relation for the full fourth order tensor $\eta_{ijkl}$.

In the second approach (used in subsection~\ref{subsection:withspin}), we consider the momentum $\rho \bm{v}$ as a vectorial quantity, enabling for the use of tensor representation theorems. We apply the flux hypothesis directly on the fourth order tensor $\eta_{ijkl}$ to recover Green-Kubo relations. 

\subsection{Viscosity for Systems with No Internal Spin}\label{subsection:nospin}
We now apply the flux hypothesis to recover Green-Kubo relations for viscosity $\bm{\eta}^{(4)}$ in 2d isotropic systems as presented in \cite{Epstein2020}. We first present Green-Kubo relations for the case of systems with no internal spin. In this case, the stress tensor $\bm{\sigma}$ is linearly dependent on the velocity gradient $\bm{\nabla}\bm{v}$ alone as
\begin{equation} \label{eq-SI:newtonian-no-spin}
    \sigma_{ij} = \eta_{ijkl}\partial_l v_{k} \,,
\end{equation}
with the corresponding balance law of linear momentum,
\begin{equation}\label{eq-SI:lin-mom-bal-nofour}
    \rho\dot{v}_i = \partial_j \sigma_{ij} \,.
\end{equation}
This balance law holds in both the macroscopic sense and  instantaneously for any microscopic realization. 

Considering \textit{microscopic fluctuations} from a spatially homogeneous steady state with density $\rho_0$ and no average bulk velocity, we may write the linearized Fourier-transformed balance law as
\begin{equation}\label{eq-SI:lin-mom-bal}
    \rho_0\partial_t\hat{ v}_i^{\bq} = \mathrm{i}q_j \hat{\sigma}_{ij}^{\bq} \,.
\end{equation}
Using matrix-vector products, the constitutive law \eqref{eq-SI:newtonian-no-spin} may be translated to its Fourier counterpart and then put into the form of Eq.~\eqref{eq:constitutive-fourier} as
\begin{equation}
    \bm{t}_x^{\bq}:= 
    \begin{bmatrix}
        \sigma_{xx}^{\bq} \\ \sigma_{xy}^{\bq} 
    \end{bmatrix} =  
    \begin{bmatrix}
        \eta_{xxxx} & \eta_{xxxy}\\
        \eta_{xyxx} & \eta_{xyxy}
    \end{bmatrix} 
    \begin{bmatrix}
        \mathrm{i}q_x \\ \mathrm{i}q_y
    \end{bmatrix} v_x^{\bq} + 
    \begin{bmatrix}
        \eta_{xxyx} & \eta_{xxyy}\\
        \eta_{xyyx} & \eta_{xyyy}
    \end{bmatrix} 
    \begin{bmatrix}
        \mathrm{i}q_x \\ \mathrm{i}q_y
    \end{bmatrix} v_y^{\bq} \,,
\end{equation}

\begin{equation}
    \bm{t}_y^{\bq}:=
    \begin{bmatrix}
        \sigma_{yx}^{\bq} \\ \sigma_{yy}^{\bq} 
    \end{bmatrix} =  
    \begin{bmatrix}
        \eta_{yxxx} & \eta_{yxxy}\\
        \eta_{yyxx} & \eta_{yyxy}
    \end{bmatrix} 
    \begin{bmatrix}
        \mathrm{i}q_x \\ \mathrm{i}q_y
    \end{bmatrix} v_x^{\bq} + 
    \begin{bmatrix}
        \eta_{yxyx} & \eta_{yxyy}\\
        \eta_{yyyx} & \eta_{yyyy}
    \end{bmatrix} 
    \begin{bmatrix}
        \mathrm{i}q_x \\ \mathrm{i}q_y
    \end{bmatrix} v_y^{\bq}   \,.
\end{equation}
Here, we define the fluxes $\bm{t}_i:=[\sigma_{ix}, \sigma_{iy}]^\intercal$ and define an unraveled matrix-analog to the 4th order viscosity tensor, 
\begin{equation}
    \mathbf{H}_{ij}:=
    \begin{bmatrix}
        \eta_{ixjx} & \eta_{ixjy}\\
        \eta_{iyjx} & \eta_{iyjy}
    \end{bmatrix} \,.
\end{equation}
Thus, mirroring Eq.~\eqref{eq:constitutive-fourier} we obtain
\begin{equation}
    \bm{t}_i^{\bq} = \sum_{k\in\{x,y\}}\mathbf{H}_{ik}\cdot\ibq v^{\bq}_k \,.
\end{equation}
We see that the quantities $\bm{t}_i$, $\mathbf{H}_{ij}$, and $v_i^{\bq}$ are in direct analogy to $\bm{J}_i$,  $\mathbf{M}_{ij}$, and $A_i^{\bq}$, respectively, in the main text. Accordingly, we may apply the flux hypothesis to obtain Green-Kubo relations for $\mathbf{H}_{ij}$, which from Eq.~\eqref{eq:Green-Kubo} are
\begin{equation} \label{eq-SI:H-eqn}
    \mathbf{H}_{ij}(\bq,\Delta t) = \sum_{k\in\{x,y\}}\mathbf{K}_{ik}(\bq,\Delta t)s_{kj}(\bq) \ ,
\end{equation}
where $\mathbf{K}_{ij}$ and $s_{ij}$ are defined in direct analogy to $\mathbf{L}_{ij}$ and $g_{ij}$, respectively, in the main text as follows:
\begin{equation}
    \mathbf{K}_{ij}(\bq,\Delta t):=\int_0^{\Delta t}dt\ \langle\hat{\bm{t}}^{\bq}_i(t)\hat{\bm{t}}^{-\bq}_j(0)\rangle = \int_0^{\Delta t}dt
    \begin{bmatrix}
         \langle \hat{\sigma}_{ix}^{\bq}(t) \hat{\sigma}_{jx}^{-\bq}(0)\rangle &  \langle \hat{\sigma}_{ix}^{\bq}(t) \hat{\sigma}_{jy}^{-\bq}(0)\rangle\\
        \langle \hat{\sigma}_{iy}^{\bq}(t) \hat{\sigma}_{jx}^{-\bq}(0)\rangle & \langle \hat{\sigma}_{iy}^{\bq}(t) \hat{\sigma}_{jy}^{-\bq}(0)\rangle
    \end{bmatrix} \,,
\end{equation}
\begin{equation}
    [\mathbf{s}^{-1}(\bq)]_{ij}:=\rho_0\langle\hat{v}_i^{\bq} \hat{v}_j^{-\bq}\rangle \,.
\end{equation}
Here, we note that $\mathbf{s}$ possesses an extra factor of the average mass density $\rho_0$ when compared to $\mathbf{g}$. 

For \textit{isotropic systems}, the static velocity correlator is diagonal to leading order (see SI-IV of Ref.~\cite{Epstein2020})
\begin{equation}
    \lim_{\bq\rightarrow\bm{0}}[\mathbf{s}^{-1}(\bq)]_{ij}=\rho_0\mu\delta_{ij}  \ ,
\end{equation}
where we have defined
\begin{equation}\label{eq:SI-mu}
    \mu\delta_{ij} := \lim_{\bq\rightarrow\bm{0}}\langle \hat{v}_i^{\bq} \hat{v}_j^{-\bq}\rangle \,.
\end{equation}
Substituting the definitions of $\mathbf{H}_{ij}$, $\mathbf{K}_{ij}$, and $\mathbf{s}$ into Eq.~\eqref{eq-SI:H-eqn}, we obtain the Green-Kubo relations for the viscosity tensor:
\begin{equation}\label{eq-SI:pre-eta}
    \begin{bmatrix}
        \eta_{ixjx} & \eta_{ixjy}\\
        \eta_{iyjx} & \eta_{iyjy}
    \end{bmatrix} = \lim_{\bq\rightarrow\bm{0}}
    \sum_{k\in\{x,y\}}\int_0^{\Delta t}dt
    \begin{bmatrix}
         \langle \hat{\sigma}_{ix}^{\bq}(t) \hat{\sigma}_{kx}^{-\bq}(0)\rangle &  \langle \hat{\sigma}_{ix}^{\bq}(t) \hat{\sigma}_{ky}^{-\bq}(0)\rangle\\
        \langle\hat{\sigma}_{iy}^{\bq}(t) \hat{\sigma}_{kx}^{-\bq}(0)\rangle & \langle \hat{\sigma}_{iy}^{\bq}(t) \hat{\sigma}_{ky}^{-\bq}(0)\rangle
    \end{bmatrix} \frac{1}{\rho_0\mu}\delta_{kj} \,.
\end{equation}
Equation~\eqref{eq-SI:pre-eta} can be reduced to 
\begin{equation}
    \begin{bmatrix}
        \eta_{ixjx} & \eta_{ixjy}\\
        \eta_{iyjx} & \eta_{iyjy}
    \end{bmatrix} = \frac{1}{\rho_0\mu }\lim_{\bq\rightarrow\bm{0}}\int_0^{\Delta t}dt
    \begin{bmatrix}
         \langle \hat{\sigma}_{ix}^{\bq}(t) \hat{\sigma}_{jx}^{-\bq}(0)\rangle &  \langle \hat{\sigma}_{ix}^{\bq}(t) \hat{\sigma}_{jy}^{-\bq}(0)\rangle\\
        \langle\hat{\sigma}_{iy}^{\bq}(t) \hat{\sigma}_{jx}^{-\bq}(0)\rangle & \langle \hat{\sigma}_{iy}^{\bq}(t) \hat{\sigma}_{jy}^{-\bq}(0)\rangle
    \end{bmatrix} \,.
\end{equation}
Finally, reorganizing $\mathbf{H}_{ij}$ into a 4th-order tensor $\eta_{ijkl}$, we write the Green-Kubo relations for the full viscosity tensors as
\begin{equation} \label{eq-SI:visc-no-spin}
    \eta_{ijkl} = \frac{1}{\rho_0\mu }\lim_{\bq\rightarrow\bm{0}}\int_0^{\Delta t}dt\ \langle \hat{\sigma}_{ij}^{\bq}(t) \hat{\sigma}_{kl}^{-\bq}(0)\rangle \,,
\end{equation}
with $\rho_0 \mu = Vk_{\text{B}}T_{\text{eff}}$ playing the role of effective temperature for non-equilibrium steady states. 

For 2d isotropic systems, the viscosity tensor can be further reduced from 16 to 6 coefficients through the orthogonal basis $s_{ijkl}^{(\alpha)}$ as outlined in Eq.~(8) and Table I of Ref.~\cite{Epstein2020}, where
\begin{equation}\label{eq-SI:visc-basis}
    \eta_{ijkl}=\sum_{\alpha=1}^6 \lambda_\alpha s_{ijkl}^{(\alpha)} = \lambda_1\delta_{ij}\delta_{kl} + \lambda_2(\delta_{ik}\delta_{jl} + \delta_{il}\delta_{jk} - \delta_{ij}\delta_{kl})+ \lambda_3\epsilon_{ij}\epsilon_{kl} + \lambda_4(\epsilon_{ik}\delta_{jl} + \epsilon_{jl}\delta_{ik}) + 2\lambda_5\epsilon_{ij}\delta_{kl} - 2\lambda_6\delta_{ij}\epsilon_{kl} \,,
\end{equation}
and $\{\lambda_\alpha\}$ are the relevant transport coefficients. The first three viscosities are even under parity symmetry: $\lambda_1$ and $\lambda_2$ are respectively the typical bulk and shear viscosities, and $\lambda_3$ is the rotational viscosity. The latter three viscosities are odd under parity symmetry and may arise in chiral systems; $\lambda_4$ is the ``odd'' viscosity coupling shear along one axis with normal stress along an orthogonal axis, $\lambda_5$ creates a torque in response to compression, and lastly $\lambda_6$ creates a normal stress in response to vorticity. 

Using Eq.~\eqref{eq-SI:visc-basis} in conjunction with Eq.~\eqref{eq-SI:visc-no-spin}, we obtain the following Green-Kubo relations for each of the six viscosities in 2d: 
\begin{equation}
    \lambda_1 = \frac{1}{4\rho_0\mu}\delta_{ij}\delta_{kl}\lim_{\bq\rightarrow\bm{0}}\int_0^{\Delta t}dt \ \langle \hat{\sigma}_{ij}^{\bq}(t) \hat{\sigma}_{kl}^{-\bq}(0)\rangle\,,
\end{equation}
\begin{equation}
    \lambda_2 = \frac{1}{8\rho_0\mu}(\delta_{ik}\delta_{jl} + \delta_{il}\delta_{jk} - \delta_{ij}\delta_{kl})\lim_{\bq\rightarrow\bm{0}}\int_0^{\Delta t}dt \ \langle \hat{\sigma}_{ij}^{\bq}(t) \hat{\sigma}_{kl}^{-\bq}(0)\rangle\,,
\end{equation}
\begin{equation}
    \lambda_3= \frac{1}{4\rho_0\mu}\epsilon_{ij}\epsilon_{kl}\lim_{\bq\rightarrow\bm{0}}\int_0^{\Delta t}dt \ \langle \hat{\sigma}_{ij}^{\bq}(t) \hat{\sigma}_{kl}^{-\bq}(0)\rangle\,,
\end{equation}
\begin{equation}
    \lambda_4 = \frac{1}{8\rho_0\mu}(\epsilon_{ik}\delta_{jl}+\epsilon_{jl}\delta_{ik})\lim_{\bq\rightarrow\bm{0}}\int_0^{\Delta t}dt \ \langle \hat{\sigma}_{ij}^{\bq}(t) \hat{\sigma}_{kl}^{-\bq}(0)\rangle\,,
\end{equation}
\begin{equation}
    \lambda_5 = \frac{1}{8\rho_0\mu}\epsilon_{ij}\delta_{kl}\lim_{\bq\rightarrow\bm{0}}\int_0^{\Delta t}dt \ \langle \hat{\sigma}_{ij}^{\bq}(t) \hat{\sigma}_{kl}^{-\bq}(0)\rangle\,,
\end{equation}
\begin{equation}
    \lambda_6 = -\frac{1}{8\rho_0\mu}\delta_{ij}\epsilon_{kl} \lim_{\bq\rightarrow\bm{0}}\int_0^{\Delta t}dt \ \langle \hat{\sigma}_{ij}^{\bq}(t) \hat{\sigma}_{kl}^{-\bq}(0)\rangle   \,.
\end{equation}
Unlike previous work applying the Onsager regression hypothesis to this context~\cite{Epstein2020}, the flux hypothesis holds the utility of isolating an explicit independent expression for each viscosity coefficient.

\subsection{Viscosity for Systems with Internal Spin}\label{subsection:withspin}
We now apply the flux hypothesis to develop Green-Kubo relations for viscosities in 2d isotropic systems with internal spin. Here, we showcase how to apply the flux hypothesis directly on higher order tensors. Allowing the stress tensor $\sigma_{ij}$ to linearly depend on the velocity gradient $\partial_j v_{i}$ and the spin field $m$ results in the constitutive law (see Eq. (6) in Ref.~\cite{Epstein2020}),
\begin{equation}\label{eq-SI:visc-const-law}
    \sigma_{ij} = \eta_{ijkl}\partial_l v_{k}+\gamma_{ij}m \ ,
\end{equation}
where $\eta_{ijkl}$ is the standard viscosity tensor and $\gamma_{ij}$ is another viscous transport tensor characterizing the stress response to the spin angular momentum. In the Fourier representation, Eq.~\eqref{eq-SI:visc-const-law} becomes
\begin{equation}\label{eq-SI:visc-const-law-four}
    \sigma_{ij}^{\bq} = \eta_{ijkl}\mathrm{i}q_l v_{k}^{\bq}+\gamma_{ij}m^{\bq} \,.
\end{equation}
While the linear momentum balance remains the same as Eq.~\eqref{eq-SI:lin-mom-bal-nofour}, the spin angular momentum $m$ follows the balance law 
\begin{equation}
    \rho\dot{m} = \partial_i C_{i}-\epsilon_{ij} \sigma_{ij}\,,
\end{equation}
with couple stress tensor $C_{i}$~\cite{Epstein2020}. We may once again consider \textit{microscopic fluctuations} from a spatially homogeneous steady state with density $\rho_0$ and no average bulk velocity to obtain the linearized Fourier-transformed balance laws as Eq.~\eqref{eq-SI:lin-mom-bal} and 
\begin{equation}\label{eq-SI:spin-bal}
    \rho_0\partial_t \hatm^{\bq} =  \mathrm{i}q_i\hat{C}^{\bq} _{i}-\epsilon_{ij} \hat{\sigma}^{\bq}_{ij} \,.
\end{equation}

Given the multiplicity of transport coefficients entering via tensors of different order, it is convenient to apply the flux hypothesis directly on the constitutive law via Eq.~\eqref{eq:flux-hypothesis-condition}. To that end, given Eq.~\eqref{eq-SI:visc-const-law-four}, the flux hypothesis states the fluctuations in stress $\hat{\sigma}_{ij}^{\bq}$, velocity $\tilde{v}_i^{\bq}$, and internal spin $\tilde{m}^{\bq}$ are related by
\begin{equation} \label{eq-fh-og-visc}
    \langle \hat{\sigma}_{ij}^{\bq}(\Delta t)\rangle_{\{\hat{v}_k^{\bq}(0)=\tilde{v}_k^{\bq}, \hat{m}^{\bq}(0)=\tilde{m}^{\bq}\}} = \eta_{ijkl}\mathrm{i}q_l \tilde{v}^{\bq}_k+\gamma_{ij} \tilde{m}^{\bq} \,.
\end{equation}
We now multiply by the initial velocity fluctuation $\tilde{v}_m^{-\bq}$, assume $\{\tilde{v}_j^{\bq}\}$ and $\{\tilde{m}^{\bq}\}$ respectively follow the steady-state distribution of the fluctuating variables $\{\hat{v}_j^{\bq}\}$ and $\{\hat{m}^{\bq}\}$, and then take a full ensemble average to obtain
\begin{equation}\label{eq-SI:stress-v}
   \langle \hat{\sigma}_{ij}^{\bq}(\Delta t) \hat{v}^{-\bq}_m(0)\rangle = \eta_{ijkl}\mathrm{i}q_l\langle \hat{v}^{\bq}_k(0) \hat{v}^{-\bq}_m(0)\rangle+\gamma_{ij}\langle\hat{m}^{\bq}(0) \hat{v}^{-\bq}_m(0)\rangle \,.
\end{equation}
Assuming there exist no static correlations between the velocity fluctuations and the stress fluctuations, the left-hand side term may be written using stationarity as 
\begin{align}\label{eq-SI:stress-v-station}
    \begin{split}
        \langle \hat{\sigma}_{ij}^{\bq}(\Delta t) \hv^{-\bq}_m(0)\rangle &=  \langle \hat{\sigma}_{ij}^{\bq}(0) \hv^{-\bq}_m(0)\rangle + \int_0^{\Delta t}\frac{\partial}{\partial t} \langle\hat{\sigma}_{ij}^{\bq}(t) \hv^{-\bq}_m(0)\rangle dt
        \\
        &=\int_0^{\Delta t} \frac{\partial}{\partial t}\langle\hat{\sigma}_{ij}^{\bq}(t) \hv^{-\bq}_m(0)\rangle dt
        \\
        &=\int_0^{\Delta t} \frac{\partial}{\partial t}\langle\hat{\sigma}_{ij}^{\bq}(0) \hv^{-\bq}_m(-t)\rangle dt
        \\
        &=-\int_0^{\Delta t} \langle\hat{\sigma}_{ij}^{\bq}(0)\frac{\partial}{\partial t'} \hv^{-\bq}_m(t')\rangle\bigg|_{t'=-t}dt \\
        &=-\int_0^{\Delta t} \langle\hat{\sigma}_{ij}^{\bq}(t)\frac{\partial}{\partial t'} \hv^{-\bq}_m(t')\rangle\bigg|_{t'=0}dt \,.
    \end{split}
\end{align}
Utilizing the balance of linear momentum in Eq.~\eqref{eq-SI:lin-mom-bal} with average steady-state mass density $\rho_0$, we obtain from Eqs.~\eqref{eq-SI:stress-v} and~\eqref{eq-SI:stress-v-station} 
\begin{equation} \label{eq-fh-visc}
    \frac{\mathrm{i}q_l}{\rho_0}\int_0^{\Delta t}dt \ \langle \hat{\sigma}_{ij}^{\bq}(t) \hat{\sigma}_{ml}^{-\bq}(0)\rangle = \eta_{ijkl}\mathrm{i}q_l\langle \hv^{\bq}_k \hv^{-\bq}_m\rangle+\gamma_{ij}\langle  \hatm^{\bq} \hv^{-\bq}_m\rangle \,.
\end{equation}

It is convenient to decouple $\gamma_{ij}$ from $\eta_{ijkl}$ in Eq.~\eqref{eq-fh-visc}. To this end, we may also multiply Eq.~\eqref{eq-fh-og-visc} by the initial spin fluctuation $\tilde{m}^{-\bq}$ and take a full ensemble average to yield
\begin{equation}\label{eq-SI:stress-m}
   \langle \hat{\sigma}_{ij}^{\bq}(\Delta t) \hatm^{-\bq}(0)\rangle = \eta_{ijkl}\mathrm{i}q_l\langle \hv^{\bq}_k(0) \hatm^{-\bq}(0)\rangle+\gamma_{ij}\langle \hatm^{\bq}(0) \hatm^{-\bq}(0)\rangle \,.
\end{equation}
Again, assuming there exist no static correlations in stress fluctuations and spin field fluctuations, the left-hand side of Eq.~\eqref{eq-SI:stress-m} can be expressed using stationarity as 
\begin{align}\label{eq-SI:stress-m-station}
    \begin{split}
        \left< \hat{\sigma}_{ij}^{\bq}(\Delta t) \hatm^{-\bq}(0)\right> &=  \langle \hat{\sigma}_{ij}^{\bq}(0) \hatm^{-\bq}(0)\rangle + \int_0^{\Delta t}\frac{\partial}{\partial t} \langle \hat{\sigma}_{ij}^{\bq}(t) \hatm^{-\bq}(0)\rangle dt
        \\
        &=  \int_0^{\Delta t}\frac{\partial}{\partial t} \langle \hat{\sigma}_{ij}^{\bq}(t) \hatm^{-\bq}(0)\rangle dt
        \\
        &=  \int_0^{\Delta t}\frac{\partial}{\partial t} \langle \hat{\sigma}_{ij}^{\bq}(0) \hatm^{-\bq}(-t)\rangle dt
        \\
        &=-\int_0^{\Delta t} \langle \hat{\sigma}_{ij}^{\bq}(0)\frac{\partial}{\partial t'} \hatm^{-\bq}(t')\rangle\bigg|_{t'=-t} dt 
        \\
        &=-\int_0^{\Delta t} \langle \hat{\sigma}_{ij}^{\bq}(t)\frac{\partial}{\partial t'} \hatm^{-\bq}(t')\rangle\bigg|_{t'=0} dt \,.
    \end{split}
\end{align}
Now, utilizing the balance of spin in Eq.~\eqref{eq-SI:spin-bal}, we obtain from Eqs.~\eqref{eq-SI:stress-m} and~\eqref{eq-SI:stress-m-station}
\begin{equation}\label{eq-fh-spinvisc}
    \frac{\mathrm{i}q_l}{\rho_0}\int_0^{\Delta t}dt \ \langle \hat{\sigma}_{ij}^{\bq}(t) \hat{C}_{l}^{-\bq}(0)\rangle  +\frac{\epsilon_{kl}}{\rho_0}\int_0^{\Delta t}dt \ \langle \hat{\sigma}_{ij}^{\bq}(t) \hat{\sigma}_{kl}^{-\bq}(0)\rangle  = \eta_{ijkl}\mathrm{i}q_l\langle \hv^{\bq}_k \hatm^{-\bq}\rangle+\gamma_{ij}\langle \hatm^{\bq} \hatm^{-\bq}\rangle \,.
\end{equation}

Equations~\eqref{eq-fh-visc} and~\eqref{eq-fh-spinvisc} alone may be sufficient to recover Green-Kubo relations for the tensors $\eta_{ijkl}$ and $\gamma_{ij}$ irrespective of material symmetry. In what follows, we derive Green-Kubo relations for \textit{isotropic systems}, which puts constraints on the nature of transport coefficient tensors and static correlation tensors. For isotropic systems, the static correlation tensors $\langle \hat{v}_i^{\bq} \hat{v}_j^{-\bq}\rangle, \langle\hatm^{\bq}\hat{v}_i^{-\bq}\rangle,$ and $ \langle\hat{m}^{\bq}\hat{m}^{-\bq}\rangle$ must themselves be isotropic in the limit $\bq\rightarrow\bm{0}$ as argued in SI-IV of Ref.~\cite{Epstein2020}. Thus, $\langle \hv^{\bq}_i \hv^{-\bq}_j\rangle$ must be a multiple of the Kronecker delta $\delta_{ij}$ to zeroth order in $\bq$, and the vector $\langle  \hatm^{\bq} \hv^{-\bq}_m\rangle$ must be first order in $\bq$. Accordingly, we define
\begin{equation}\label{eq-SI:mu}
    \mu\delta_{ij} := \lim_{\bq\rightarrow\bm{0}}\langle \hat{v}_i^{\bq} \hat{v}_j^{-\bq}\rangle \,,
\end{equation}
\begin{equation}\label{eq-SI:Omega}
    \Omega_{ij} := -\lim_{\bq\rightarrow\bm{0}}\frac{\partial}{\partial \mathrm{i}q_i}\langle\hatm^{\bq}\hat{v}_j^{-\bq}\rangle \,,
\end{equation}
\begin{equation}
    \pi:=\delta_{ij}\Omega_{ij}\,,
\end{equation}
\begin{equation}
    \tau:=\epsilon_{ij}\Omega_{ij}\,,
\end{equation}
\begin{equation}\label{eq-SI:nu}
    \nu := \lim_{\bq\rightarrow\bm{0}}\langle \hat{m}^{\bq} \hat{m}^{-\bq}\rangle \,.
\end{equation}
As the static correlators may be discontinuous at $\bq=\bm{0}$, they are considered at the limit $\bq\rightarrow\bm{0}$, which may or may not be strictly equal to their evaluation at $\bq=\bm{0}$. Thus, expanding around $\bq=\bm{0}^{+}$, $\langle \hv_i^{\bq} \hv_j^{-\bq}\rangle=\mu \delta_{ij}+\mathcal{O}(q)$, $\langle \hatm^{\bq} \hv_i^{-\bq}\rangle=-\langle \hv_i^{\bq} \hatm^{-\bq}\rangle=-\mathrm{i}q_l \Omega_{li}+\mathcal{O}(q^2)$, and $\langle \hatm^{\bq} \hatm^{-\bq}\rangle=\nu +\mathcal{O}(q^2)$. 

Now, considering the isotropy of the static correlators in Eq.~\eqref{eq-fh-visc} and noting the arbitrary orientation of $\bq$, we obtain the following reduced Green-Kubo relations:
\begin{equation} \label{eq-eta-viscbox}
    \rho_0\mu\eta_{ijkl}-\rho_0\Omega_{lk}\gamma_{ij} = \lim_{\bq\rightarrow\bm{0}}\int_0^{\Delta t}dt \ \langle\hat{\sigma}_{ij}^{\bq}(t) \hat{\sigma}_{kl}^{-\bq}(0)\rangle \,.
\end{equation}
Applying the same assumptions to Eq.~\eqref{eq-fh-spinvisc}, while also grouping order $\mathcal{O}(1)$ terms in the limit $\bq\rightarrow\bm{0}$ yields
\begin{equation}\label{eq-gamma-viscbox}
    \rho_0\nu\gamma_{ij} = \lim_{\bq\rightarrow\bm{0}}\epsilon_{kl}\int_0^{\Delta t}dt \ \langle \hat{\sigma}_{ij}^{\bq}(t) \hat{\sigma}_{kl}^{-\bq}(0)\rangle \,.
\end{equation}

Thus far, we have exploited the isotropy of the static correlation functions. Now, we additionally utilize the isotropy of the transport tensors $\eta_{ijkl}$ and $\gamma_{ij}$, expanding them with the bases outlined in Eqs. (7) and (8) and Table I of Ref.~\cite{Epstein2020}. The basis for $\eta_{ijkl}$ is specified in Eq.~\eqref{eq-SI:visc-basis} and the 2d isotropic basis for $\gamma_{ij}$ is specified as
\begin{equation}
    \gamma_{ij} = \sum_{\alpha=1}^2 \gamma_\alpha s_{ij}^{(\alpha)} = \gamma_1 \delta_{ij} + \gamma_2 \epsilon_{ij} \,.
\end{equation}
The transport coefficients $\gamma_1$ and $\gamma_2$ couple the spin field to a normal stress and an antisymmetric stress, respectively. Using these bases in conjunction with Eqs.~\eqref{eq-eta-viscbox} and \eqref{eq-gamma-viscbox}, we obtain the final Green-Kubo relations for each transport coefficient relating the stress tensor to the velocity gradient and spin field, 
\begin{equation}
    \lambda_1 = \frac{1}{4\rho_0\mu}\left(\delta_{ij}\delta_{kl}+\frac{\pi}{\nu}\delta_{ij}\epsilon_{kl}\right)\lim_{\bq\rightarrow\bm{0}}\int_0^{\Delta t}dt\left< \hat{\sigma}_{ij}^{\bq}(t) \hat{\sigma}_{kl}^{-\bq}(0)\right>\,,
\end{equation}
\begin{equation}
    \lambda_2 = \frac{1}{8\rho_0\mu}(\delta_{ik}\delta_{jl} + \delta_{il}\delta_{jk} - \delta_{ij}\delta_{kl})\lim_{\bq\rightarrow\bm{0}}\int_0^{\Delta t}dt\left< \hat{\sigma}_{ij}^{\bq}(t) \hat{\sigma}_{kl}^{-\bq}(0)\right>\,,
\end{equation}
\begin{equation}
    \lambda_3= \frac{1}{4\rho_0\mu}\left(1-\frac{\tau}{\nu}\right)\epsilon_{ij}\epsilon_{kl}\lim_{\bq\rightarrow\bm{0}}\int_0^{\Delta t}dt\left< \hat{\sigma}_{ij}^{\bq}(t) \hat{\sigma}_{kl}^{-\bq}(0)\right>\,,
\end{equation}
\begin{equation}
    \lambda_4 = \frac{1}{8\rho_0\mu}(\epsilon_{ik}\delta_{jl}+\epsilon_{jl}\delta_{ik})\lim_{\bq\rightarrow\bm{0}}\int_0^{\Delta t}dt\left< \hat{\sigma}_{ij}^{\bq}(t) \hat{\sigma}_{kl}^{-\bq}(0)\right>\,,
\end{equation}
\begin{equation}
    \lambda_5 = \frac{1}{8\rho_0\mu}\left(\epsilon_{ij}\delta_{kl}+\frac{\pi}{\nu}\epsilon_{ij}\epsilon_{kl}\right)\lim_{\bq\rightarrow\bm{0}}\int_0^{\Delta t}dt\left< \hat{\sigma}_{ij}^{\bq}(t) \hat{\sigma}_{kl}^{-\bq}(0)\right>\,,
\end{equation}
\begin{equation}
    \lambda_6 = -\frac{1}{8\rho_0\mu}\left(1-\frac{\tau}{\nu}\right)\delta_{ij}\epsilon_{kl} \lim_{\bq\rightarrow\bm{0}}\int_0^{\Delta t}dt\left< \hat{\sigma}_{ij}^{\bq}(t) \hat{\sigma}_{kl}^{-\bq}(0)\right>\,,
\end{equation}
\begin{equation}
    \gamma_1 = \frac{1}{2\rho_0\nu}\delta_{ij} \epsilon_{kl}\lim_{\bq\rightarrow\bm{0}}\int_0^{\Delta t}dt\left< \hat{\sigma}_{ij}^{\bq}(t) \hat{\sigma}_{kl}^{-\bq}(0)\right> \,,
\end{equation}
\begin{equation}
    \gamma_2 = \frac{1}{2\rho_0\nu}\epsilon_{ij} \epsilon_{kl}\lim_{\bq\rightarrow\bm{0}}\int_0^{\Delta t}dt\left< \hat{\sigma}_{ij}^{\bq}(t) \hat{\sigma}_{kl}^{-\bq}(0)\right> \,.
\end{equation}
Once again, using the flux hypothesis, we isolate an independent expression for each transport coefficient, which would be unattainable using the ORH without additional assumptions.

The above fully separated Green-Kubo relations reveal equivalencies between two pairs of transport coefficients. Namely, the antisymmetric stress originating from vorticity (governed by $\lambda_3$) is related to the antisymmetric stress originating from the spin field (governed by $\gamma_2$). Furthermore, the normal stress originating from vorticity (governed by $\lambda_6$) is related to the normal stress originating from the spin field (governed by $\gamma_1$). These findings can be expressed in the following equalities, 
\begin{equation}
    -\frac{4\lambda_6}{\gamma_1} =\frac{2\lambda_3}{\gamma_2} =\frac{(\nu-\tau)}{\mu} \,.
\end{equation}
While the second equality has been shown in \cite{Epstein2020} via the regression hypothesis, the flux hypothesis reveals an additional connection between $\lambda_6$ and $\gamma_1$. Note in equilibrium due to equipartioning, there exist no static cross-correlations between spin and velocity, making $\tau=0$. Furthermore, in equilibrium $\nu/\mu = I/\rho_0$, with moment of inertia density $I$. Thus in equilibrium systems, equivalencies can be made between transport coefficients relating a stress to vorticity and with those relating that same stress to the spin field. Stresses then stem from the difference of the vorticity and intrinsic spin field.

In summary, we demonstrated how to apply the flux hypothesis to higher-order tensors directly, which leads to Green-Kubo relations for individual transport coefficients. 
This is in contrast to the work presented in~\cite{Epstein2020}, which utilizes the Onsager regression hypothesis and yields fewer equations, making it impractical and perhaps impossible to generally isolate expressions for each transport coefficient without additional assumptions, thus showing the utility of the flux hypothesis. 
The flux hypothesis allows for all information pertinent to a transport coefficient to be utilized, whether that coefficient affects temporal relaxation or not.

\subsection{Thermal Conductivity}\label{subsection:heat}
We now apply the flux hypothesis to derive a Green-Kubo relation for the thermal conductivity tensor for 2d isotropic systems, including both the even and odd thermal conductivities~\cite{Fruchart2022oddideal}. Neglecting body heating and viscous dissipation, the balance law for the specific internal energy $u$ is
\begin{equation}\label{eq-SI:energy-balance}
    \rho\dot{u} = -\bm{\nabla}\cdot\bm{Q} \,,
\end{equation}
which holds for each microscopic realization with $\bm{Q}$ being the conductive heat flux vector. Considering \textit{microscopic fluctuations} from a spatially homogeneous steady state with no average bulk velocity, we may write the linearized Fourier-transformed balance law as
\begin{equation}
    \rho_0\partial_t\hat{u}^{\bq} = -\ibq\cdot \hat{\bm{Q}}^{\bq} \,.
\end{equation}
To close this differential equation, we necessitate a constitutive law relating the heat flux to a field variable. Assuming Fourier's Law, which states that the heat flux $\bm{Q}$ is linearly related to the temperature gradient $\bm{\nabla}T$, we write
\begin{equation}
    \bm{Q} = -\bm{\kappa}\cdot\bm{\nabla} T\,,
\end{equation}
with $\bm{\kappa}$ being the second order thermal conductivity tensor.
Further assuming that temperature and energy are related, we may express Fourier's law as 
\begin{equation}\label{eq-SI:fourier-law}
    \bm{Q} = -\bm{\kappa}\cdot\frac{dT}{du}\bm{\nabla} u \,,
\end{equation}
with $\frac{dT}{du}$ being the inverse specific heat capacity, which we will take to be constant.

The quantities $\bm{Q}$, $u$, and $\bm{\kappa}$ are in direct analogy to the quantities $\bm{J}_i$, $A_i$, and $\mathbf{M}_{ij}$ in the main text. Thus, mirroring Eq.~\eqref{eq:Green-Kubo}, we have
\begin{equation} 
    \bm{\kappa} = \mathbf{G}(\bq,\Delta t)c(\bq) \ ,
\end{equation}
where $\mathbf{G}$ and $c$ are 
\begin{equation}
    \mathbf{G}(\bq,\Delta t)=\int_0^{\Delta t}dt\langle\hat{\bm{Q}}^{\bq}(t)\hat{\bm{Q}}^{-\bq}(0)\rangle \,,
\end{equation}
\begin{equation}
    c^{-1}(\bq)=\rho_0\frac{dT}{du}\langle \hat{u}^{\bq} \hat{u}^{-\bq}\rangle \,.
\end{equation}
Thus, the Green-Kubo relation for the thermal conductivity tensor becomes
\begin{equation}\label{eq-SI:gk-thermal}
     \bm{\kappa} =\lim_{\bq\rightarrow\bm{0}}\frac{1}{\rho_0\frac{d T}{d u}\left< \hat{u}^{\bq} \hat{u}^{-\bq}\right>}\int_0^{\Delta t}dt\langle\hat{\bm{Q}}^{\bq}(t)\hat{\bm{Q}}^{-\bq}(0)\rangle \,.
\end{equation}

For 2d isotropic systems, we may expand the thermal conductivity tensor using the representation theorem in SI-I of Ref.~\cite{Epstein2020} as
\begin{equation}\label{eq-SI:kappa-representation}
    \kappa_{ij} = \kappa_{\parallel}\delta_{ij} + \kappa_{\perp}\epsilon_{ij} \ ,
\end{equation}
where $\kappa_\parallel$ and $\kappa_\perp$ refer to the even and odd thermal conductivity, respectively. A non-zero odd thermal conductivity $\kappa_\perp$ induces heat fluxes orthogonal to temperature gradients. By contracting $\delta_{ij}$ or $\epsilon_{ij}$ with Eq.~\eqref{eq-SI:gk-thermal}, we obtain
\begin{equation}\label{eq-SI:gk-thermal-even}
     \kappa_{\parallel} =\lim_{\bq\rightarrow\bm{0}}\frac{1}{2\rho_0\frac{d T}{d u}\left< \hat{u}^{\bq} \hat{u}^{-\bq}\right>}\int_0^{\Delta t}dt\langle\hat{Q}_i^{\bq}(t)\hat{Q}_j^{-\bq}(0)\rangle\delta_{ij}\,,
\end{equation}
\begin{equation}\label{eq-SI:gk-thermal-odd}
     \kappa_{\perp} =\lim_{\bq\rightarrow\bm{0}}\frac{1}{2\rho_0\frac{d T}{d u}\left< \hat{u}^{\bq} \hat{u}^{-\bq}\right>}\int_0^{\Delta t}dt\langle\hat{Q}_i^{\bq}(t)\hat{Q}_j^{-\bq}(0)\rangle\epsilon_{ij} \,.
\end{equation}
Note with the substitution $\frac{du}{dT} = \rho_0 V \langle \hat{u} ^2\rangle/(k_{\text{B}}T^2)$, we recover the usual pre-factor of $V/(k_{\text{B}}T^2)$. Equation~\eqref{eq-SI:gk-thermal-even} is the standard Green-Kubo relation for the symmetric part of the thermal conductivity tensor. However, as can be seen from combining Eqs.~\eqref{eq-SI:energy-balance},~\eqref{eq-SI:fourier-law} and~\eqref{eq-SI:kappa-representation}, there exist no contributions of $\kappa_\perp$ to energy relaxation due to the divergence free nature of odd fluxes. Therefore, there exists no possibility of obtaining the Green-Kubo relations for $\kappa_\perp$ from the Onsager regression hypothesis. The flux hypothesis, which instead acts on the level of the constitutive law~\eqref{eq-SI:fourier-law}, is able to yield a Green-Kubo relation for $\kappa_\perp$, showing that both time-reversal symmetry and parity symmetry breaking is necessary to observe odd thermal fluxes.  

\end{document}